\documentclass{aastex701}

\usepackage{amsmath}
\journalinfo{Published in The Planetary Science Journal, 7, 169 (2026)}

\begin{document}

\title{Planet-wide, concentric density waves in Venus's upper atmosphere revealed through polarimetry?}

\author[orcid=0000-0003-2613-4413,sname='Gourav']{Gourav Mahapatra}
\affiliation{Faculty of Aerospace Engineering, Delft University of Technology, Delft, The Netherlands}
\email[show]{mahapatra.gourav@gmail.com}
\correspondingauthor{Gourav Mahapatra}

\author{Michiel Rodenhuis} 
\affiliation{Leiden Observatory, Leiden University, Leiden, The Netherlands}
\email{rodenhuis@strw.leidenuniv.nl}

\author{Frans Snik}
\affiliation{Leiden Observatory, Leiden University, Leiden, The Netherlands}
\email{snik@strw.leidenuniv.nl}

\author[orcid=0000-0003-3697-2971]{Daphne M. Stam}
\affiliation{Leiden Observatory, Leiden University, Leiden, The Netherlands}
\email{stam@strw.leidenuniv.nl}

\author[orcid=0000-0003-4244-3419]{Lo\"ic Rossi}
\affiliation{Faculty of Aerospace Engineering, Delft University of Technology, Delft, The Netherlands}
\email{loic.cg.rossi@gmail.com}

\author[orcid=0000-0002-1368-841X]{Christoph Keller}
\affiliation{National Solar Observatory, 3665 Discovery Dr, Boulder, CO 80303, USA}
\affiliation{Leiden Observatory, Leiden University, Leiden, The Netherlands}
\email{keller@strw.leidenuniv.nl}

\begin{abstract}

We report observations of faint (10$^{-6}$), concentric, planet-wide rings in 
the polarized flux of sunlight that is reflected by Venus, obtained during a 
serendipitous, 36-minute run in 2010, with the highly-sensitive Extreme Polarimeter 
(ExPo) on the William Herschel Telescope. 
The rings appear to be centered slightly downwind of the sub-solar point, 
are visible in different filters across the visible, and are not 
obvious in the simultaneous total flux observations. 
ExPo's dual-beam exchange and double-differencing design strongly suppresses 
first-order instrumental polarization, and we could not identify an instrumental 
cause of the observed pattern.
Because ExPo was dismantled before the rings were identified in the data, 
this is the only set of observations of these rings. We are therefore 
careful in claiming the detection of a new atmospheric phenomenon on Venus.
However, numerical radiative transfer simulations show that planet-wide rings 
in polarization can arise due to density variations of 5 - 10\% in the gas above 
the clouds, consistent with a gravity-wave. Our simulations also show that 
such density variations would not show up in total flux observations. 
By presenting our observations and numerical simulations, we hope to motivate new 
polarimetric observations of Venus that could confirm or refute the presence 
of such planet-wide waves.

\end{abstract}


\section{Introduction} 

There is ample evidence for gravity wave activity on Venus. Indeed, waves with
various wavelengths have been observed through direct imaging (e.g.\ Venus 
Express and Akatsuki images) as well as through in-situ measurements \citep{piccialli2014high, Peralta2017, kasprzak1988wavelike, persson2015venus, migliorini2011oxygen}.
The properties of gravity waves depend on the atmospheric structure and 
on the thermal and dynamical interactions within the atmosphere. A recent example of planet-wide wave interaction is the bow-shaped planet-wide gravity wave detected by the Akatsuki Spacecraft just after it entered into the orbit around Venus. These waves were topographic gravity waves, generated by near-surface flows impinging on mountains and propagating upward. Since these waves possess momentum in the direction opposite to Venus' super-rotation, their dissipation at high altitudes exerts deceleration on the mean flow velocity \citep{kitahara2019mountain}. Venus' middle and upper atmosphere has long been expected to support gravity waves 
with wavelengths ranging from 100 to 600~km \citep{kasprzak1988wavelike}. It has been conjectured that wave drag and turbulence because of such internal gravity waves propagating in Venus' upper atmosphere is critical to understanding its super-rotation \citep{alexander1992mechanism, alexander1993local}. Detecting 
and characterizing them both on Venus' dayside and nightside would 
improve our understanding of our still enigmatic neighboring planet's dynamic atmosphere.

The density variations in the thin atmosphere above the clouds will be quite 
small, and very challenging to detect with flux measurements of e.g.\
reflected sunlight, in particular because the bright, underlying clouds 
will completely dominate the signal. Polarimetry could facilitate the detection of subtle atmospheric density changes in the upper atmosphere of Venus because of its higher sensitivity through the process of Rayleigh scattering by the gas molecules.

This article presents ground-based imaging polarimetry of Venus at various visible wavelengths using the Extreme Polarimeter (ExPo) instrument \citep{2007lyot.confQ..43R, canovas2011data}. We investigate an apparent ring-like polarization pattern in a unique serendipitous data set and explore whether it could plausibly be explained by gas-density variations above the cloud tops. ExPo is an experimental instrument developed to provide high-contrast imaging polarimetry, with the primary objective of observing weak sources, such as exoplanets, in reflected (and thus linearly polarized) starlight at small angular separations from their host stars. The instrument is highly sensitive to linearly polarized light, capable of achieving a sensitivity of 10$^{-4}$ provided sufficient photons are available \citep{rodenhuis2008extreme}. Because the observations are limited to a single serendipitous data set and cannot be repeated with the same instrument, our goal is not to claim a definitive detection of a new Venusian wave phenomenon, but to assess whether the observed pattern is physically plausible and worthy of targeted future observations.

In Sect.~\ref{observations}, we describe the observations. In Sect.~\ref{methods}, we outline the observational and numerical methods employed in this study, including definitions of polarized light observations (Sect.~\ref{definitions}), a description of the instrument (Sect.~\ref{instrument}), and the radiative transfer algorithm (Sect.~\ref{rtalgorithm}). Section~\ref{observation_explanation} provides a discussion of the observations, while Section~\ref{results_discussion} examines the observations in conjunction with the results of the radiative transfer model and their implications.

\section{The observations}
\label{observations}

\begin{figure}[t!]
\centering
\includegraphics[width=0.9\linewidth]{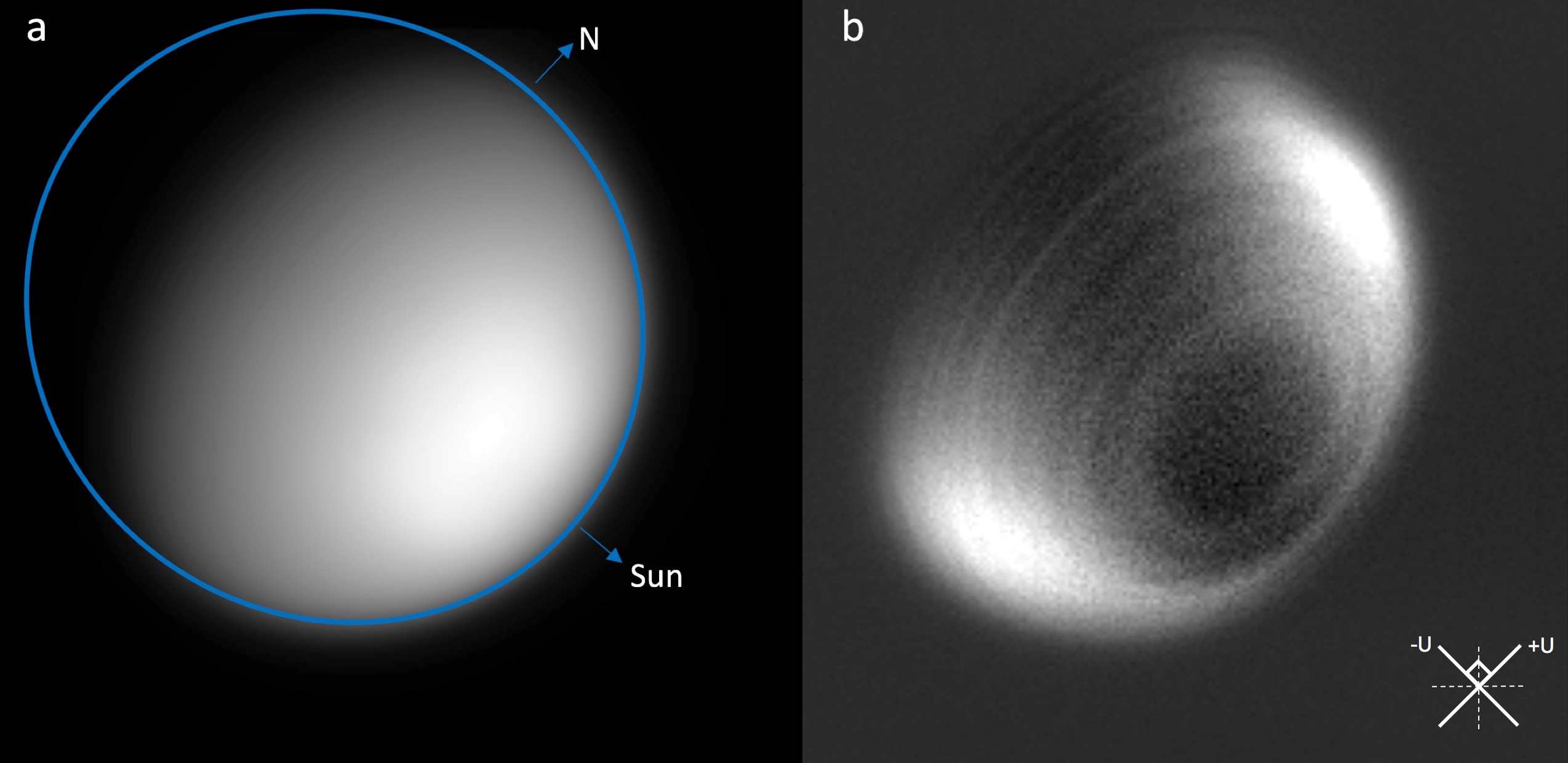}
\caption{Venus as observed in total flux $F$ (left) and polarized flux $U$ 
        (right)
        through the H-alpha continuum filter (see Table~\ref{filterspecs}).
        Due to Venus's orientation, the planetary scattering plane makes 
        an angle of about 45$^\circ$ with ExPo's optical plane 
        (the two orthogonal directions along which ExPo measured $U$ are 
        indicated below the image). 
        The spatial resolution is approximately 775~km at the sub-observer point
        on Venus' disk.}
\label{fig1}
\end{figure}

The Venus images we present here are truly serendipitous: 
they were captured with the Extreme Polarimeter (ExPo) as it was installed at the Nasmyth focus of the 4.2~m William Herschel Telescope (WHT) 
of the Isaac Newton Group (ING) of telescopes on La Palma, Canary Islands, Spain, 
during a period of about 36~minutes in which we were waiting for the end of 
the astronomical evening twilight on 24~May 2010 to start observations of circumstellar disks. 
The images were taken through six optical filters present in the instrument filter wheel: H-alpha, H-alpha continuum, Na continuum, Sloan~r, Sloan~i, and Na. 
The filter specifications and the times at which they were used are listed in 
Table~\ref{filterspecs}. 
ExPo was designed to capture images of faint, linearly polarized sources
by suppressing the contributions of unpolarized light to a high degree 
\citep{rodenhuis2012extreme}. The seeing on the evening of 
the observations (May~24th, 2010), was excellent (0.8'' or better), although 
the horizon where Venus was observed during the last observations just before
setting, appeared to be hazy possibly leading to increased atmospheric seeing values. 
The absolute accuracy of these impromptu images (thus, the background
polarization level) can only be 
calibrated to 1-2\%. This, however, is no limitation for their interpretation 
because ExPo has been designed to reach a very high polarimetric sensitivity of 10$^{-4}$.

\begin{table}[b]
\caption{We observed Venus with ExPo on the WHT, La Palma, Spain,
         on 24 May 2010. During our observations, 
         Venus' phase angle was 48.7$^\circ$, and its angular diameter 12.5''
         (outside the Earth's atmosphere).
         The technical parameters of the observations are listed below, with for 
         each filter:
         the central wavelength $\lambda_0$ and width $\Delta \lambda$. 
         And for each observation: the angle $\theta$ of ExPo's 
         ferro-electric liquid crystal (FLC),
         the time, Venus' elevation and airmass,
         and the visibility of the rings in the polarized flux. Venus's angular diameter and airmass at the times of the observations
         were obtained using 
         JPL's Horizons ephemeris tool.}
\vspace*{0.5cm}
\resizebox{0.9\textwidth}{!}{
\begin{tabular}{l c c c c c c c} \hline
Filter 	& $\lambda_0$ [nm] & $\Delta \lambda$ [nm] & $\theta$ [$^\circ$]
   & Time [UTC] & Elevation [$^\circ$] & Airmass & Rings visible \\ \hline \hline
H--alpha 	& 656.3 & \hspace*{0.22cm} 1.0 	& 45.0 & 20:28:18  &  24.06 & 2.438 & yes \\
    					& 		& 		& 67.5 & 20:31:03  & 23.43 & 2.499 & yes \\
                        \hline 
H--alpha cont. 	& 647.1	& \hspace*{0.21cm} 1.0 	    & 0.0	 & 20:33:45 & 22.80 & 2.541 &  yes \\
    					& 		& 		& 22.5 & 20:39:14 & 21.75 & 2.678 & yes \\
                        & 		& 		& 45.0 & 20:41:33 & 21.34 & 2.727 & yes \\
                        & 		& 		& 67.5 & 20:43:59 & 20.71 & 2.778 & yes \\
                        \hline 
Na  & 589.4 & \hspace*{0.22cm} 5.0 	& 45.0 & 20:51:38 &  19.05 & 3.003 & yes \\
    					& 		& 		& 67.5 & 20:54:01	& 18.63 & 3.097 & yes \\
    					\hline
Na cont.  & 580.0 & \hspace*{0.22cm} 5.0 	& 45.0 & 21:03:19 & 16.77 & 3.419 & no \\
                        & 		& 		& 67.5 & 21:06:30	& 16.16 & 3.542 & no \\
                        \hline 
Sloan r  & 623.1 & 137.3 & 45.0 & 21:23:30 &  12.68 & 4.449 & no \\
                        & 		& 		& 67.5 & 21:25:39 &  12.07 & 4.586 & no \\
                        \hline 
Sloan i  & 762.5 & 152.6 & 45.0 & 21:32:21 &  10.85 & 5.139 & no \\
                        & 		& 		& 67.5 & 21:34:31 &  10.45 & 5.322 & no \\
                        \hline 
\end{tabular}}
\label{filterspecs}
\end{table}

Figure~\ref{fig1} shows Venus in total (left) and linearly polarized (right)
fluxes as imaged through the H--alpha continuum filter (see Table ~\ref{filterspecs}). 
While the total flux image shows a homogeneous planet that is brightest 
near the sub--solar region, the polarization image shows 
relatively high polarized fluxes across the North and South polar regions, 
low polarization at lower latitudes, all overlaid with what appear to be narrow, 
higher polarization rings.\footnote{Although we do not know whether all 
arcs form full rings across the 
planet, we will refer to them as 'rings.'}
The high polarization across Venus' poles
has been observed before, for example, by Orbiter Cloud Photopolarimeter (OCPP)  \citep{1979SPIE..183..299T} 
on the Pioneer Venus spacecraft \citep{kawabata1980cloud,Sato1996} and by
Spectroscopy for Investigation of Characteristics of the Atmosphere of Venus (SPICAV) on Venus Express \citep{rossi2015preliminary}. It is worth noting that the Pioneer Venus OCPP had a polarimetric sensitivity of approximately 10$^{-3}$ \citep{hunten2022venus}, which is a magnitude lower than that of ExPo. The SPICAV-IR instrument on Venus Express, while capable of deriving partial polarization from orthogonally polarized flux measurements, was not specifically calibrated for polarimetric observations. Consequently, the exact polarimetric sensitivity of SPICAV-IR remains unknown. The pattern appears roughly concentric around a region slightly downwind of the sub-solar point and extends across much of the illuminated disk. They were observed in the first three observations through the H-alpha, H-alpha continuum, and Na, but not in the later Na continuum, Sloan r, and Sloan i observations.
The typical difference between the polarized flux inside and outside the 
brightest ring in the image in Fig.~\ref{fig1}, is $5 \cdot 10^{-4}$.

\begin{figure}[t]
\centering
\includegraphics[width=0.92\linewidth]{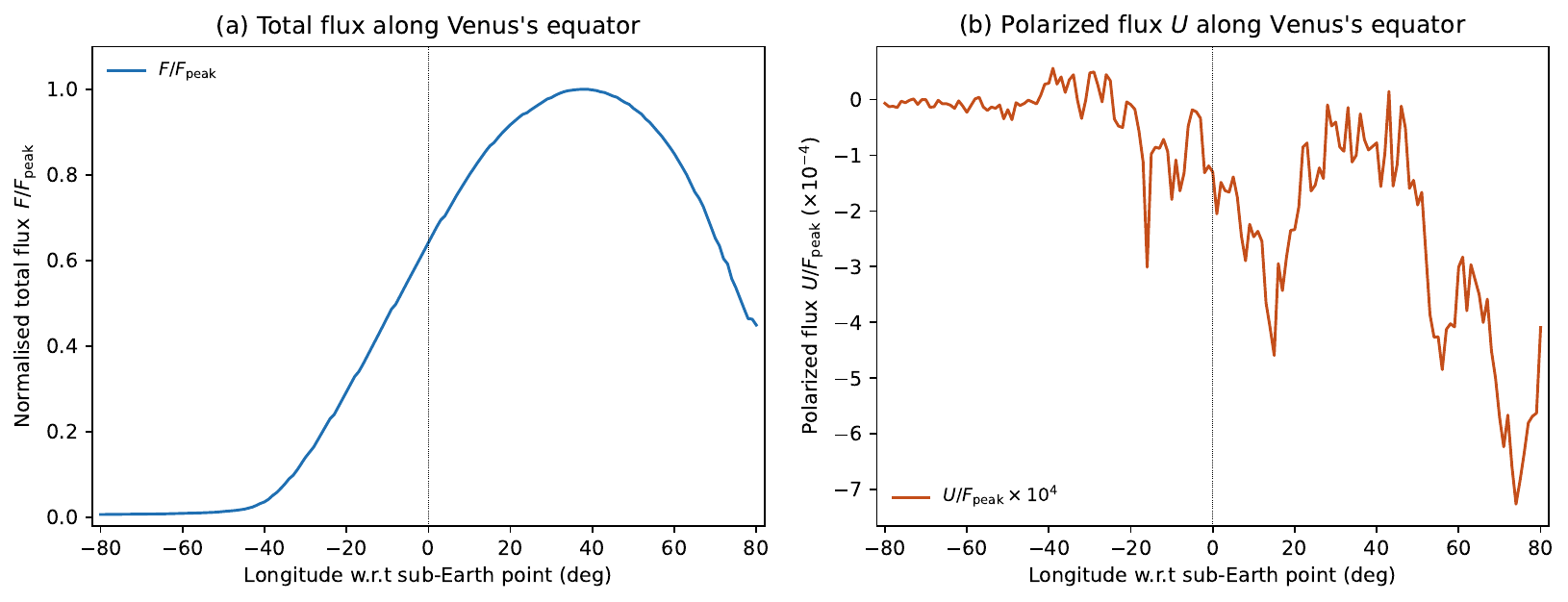}
\caption{Equatorial cut through the H--alpha continuum observation shown in Fig.~\ref{fig1}, averaged over latitudes $-5^\circ \leq \mathrm{lat} \leq +5^\circ$. Panel (a): total flux $F/F_{\rm peak}$ as a function of sub-observer longitude. Panel (b): polarized flux $U/F_{\rm peak}$ along the same strip, scaled by $10^{4}$ for readability. The total-flux cut shows no coherent ring-scale modulation, whereas the polarized-flux cut exhibits the structure corresponding to the ring-like pattern visible in the polarized image of Fig.~\ref{fig1}.}.
\label{fig:eqcut}
\end{figure}

Figure~\ref{fig:eqcut} provides a more quantitative comparison between the total and polarized flux in the same H--alpha continuum observation shown in Fig.~\ref{fig1}. We extract a one-dimensional equatorial cut averaged over the latitude band $-5^\circ \leq \mathrm{lat} \leq +5^\circ$, i.e.\ the same strip that is used later in the longitude-resolved comparison. Panel~(a) shows that the total flux is dominated by the smooth limb-darkening profile and does not show a detectable coherent ring-scale modulation. Panel~(b), by contrast, shows that the polarized flux $U$ along the same strip exhibits a clear oscillatory pattern with amplitude of order $\sim 10^{-4}\,F_{\rm peak}$ on top of the slowly varying continuum polarization.

\begin{figure}[!h]
\centering
\includegraphics[width=0.8\linewidth]{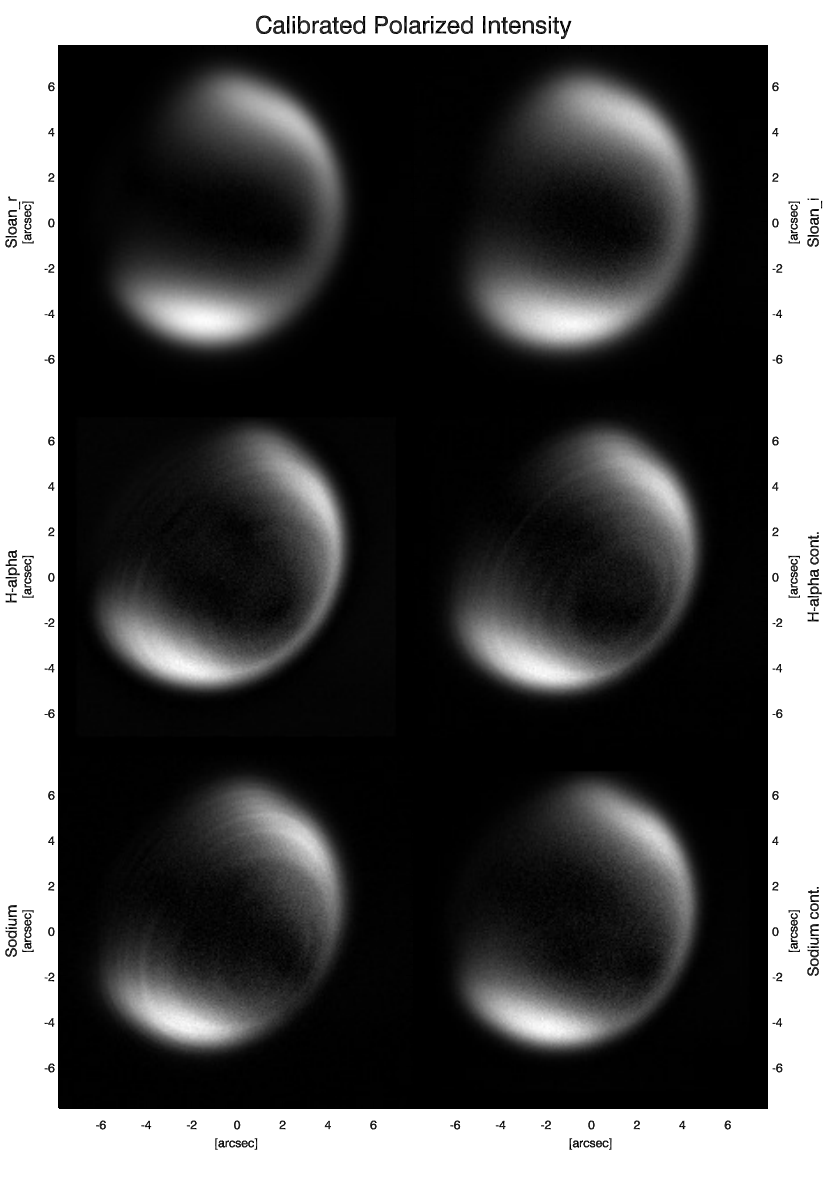}
\caption{The polarized flux ($\sqrt{Q^2+U^2}$) as measured through the six 
         filters (see Table~\ref{filterspecs}). 
         Top row: the Sloan $r$ and Sloan $i$ broadband filters. 
         Middle row: the H--alpha and H--alpha continuum filters. 
         Bottom row: the Na (sodium) and Na continuum filters. 
         The time sequence of the observations is from top left to bottom right: 5--6--1--2--3--4. 
         The ring-like pattern is apparent in the first three observations in chronological order (1,2,3) and not obvious in the last three (4,5,6). 
         Notably, the Na image at 589.4~nm still shows a clear ring pattern despite having been obtained at higher airmass than the H--alpha and H--alpha continuum observations.}
\label{fig:calPol}
\end{figure}

As detailed in Table \ref{filterspecs}, various other filters were used to observe Venus on the same day during this observation campaign. Fig.~\ref{fig:calPol} shows our six observations
of Venus in polarized flux which is defined as $\sqrt{Q^2+U^2}$. In all images, Venus shows a similar background 
polarized flux that is small in the equatorial region and high towards the
poles.

Three of the images (obtained through the H--alpha, H--alpha cont., and Na filters) show concentric ring-like patterns, while three (obtained through the Na cont., Sloan $r$, and Sloan $i$ filters) do not. The latter three observations were obtained later in the sequence (see Table~1), when Venus was lower on the horizon and the airmass was larger. However, the visibility of the features does not show a simple monotonic dependence on airmass alone: the Na observation at 589.4~nm, obtained after the H--alpha and H--alpha continuum observations and at somewhat higher airmass, still shows a particularly clear pattern. This indicates that increased airmass and reduced spatial resolution cannot by themselves explain the detect versus non-detect behavior. Instead, the observations are more likely shaped by a combination of effects, including wavelength-dependent sensitivity of the polarization signal, atmospheric seeing, and the lack of an operational atmospheric dispersion corrector in ExPo during these observations, which may have introduced some spectral smearing in the broader filters.

\begin{figure}[ht]
\centering
\includegraphics[width=0.8\linewidth]{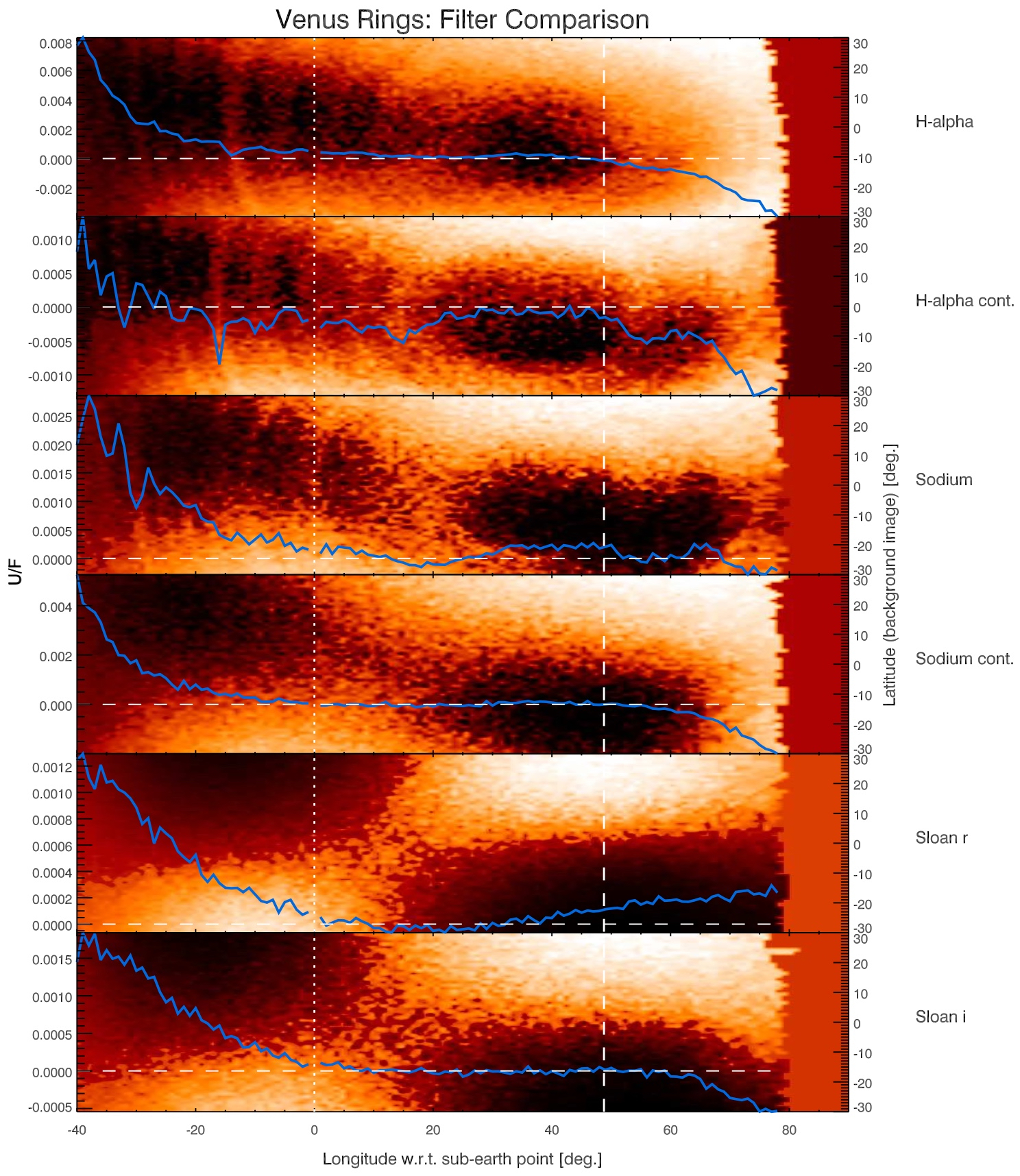}
\caption{The background (red-orange) shows the degree of polarization 
         as a function of longitude with respect to the sub--observer point 
         (horizontal axis) and the latitude from -30$^\circ$ to +30$^\circ$
         on Venus (vertical axis)
         as measured in the various filters,
         with the time of observation increasing from top to bottom
         (see Table~\ref{filterspecs}).
         The blue line shows the degree of polarization averaged over the 
         latitudes between -5$^\circ$ and +5$^\circ$ as a function of longitude
         with respect to the sub--observer point (scale on the left).
         The degree of polarization equals $U/F$, with $U$ defined with respect to the instrumental optical plane.
         Because the orientation of the planet with respect to this plane
         is almost equal to 45$^\circ$, Stokes parameter $U$ in the instrumental
         optical plane corresponds almost to Stokes parameter $Q$ in the 
         planetary scattering plane. The difference between the planetary 
         scattering plane and the instrumental optical plane is 3 to 5 degrees.
         The patterns are less visible in Stokes parameter $Q$ defined with
         respect to the instrumental optical plane (and, hence, they are much
         less visible in Stokes parameter $U$ as defined with respect to the 
         planetary scattering plane), indicating that the direction of 
         polarization of the background signal and the rings is mostly 
         parallel to the planetary scattering plane (see Fig.~\ref{fig5}).
         The vertical dotted line indicates the sub-observer or sub-earth
         longitude, and the vertical dashed line the sub-solar longitude.
         The horizontal long-dashed line indicates the zero polarization
         level for the blue lines. Note that because of the lack of 
         accurate absolute calibration, this blue line and the 
         zero level have an uncertainty of 1-2\%. The shape of the 
         continuum polarization is also influenced by the difference 
         between the planetary scattering plane and the instrumental optical
         plane.
         }
\label{fig3}
\end{figure}

Figure~\ref{fig3} shows the degree of polarization $U/F$ for the six observations,
except only for the latitudes between -30$^\circ$ and +30$^\circ$, as functions
of the longitude on the planet measured with respect to the sub-observer 
or sub-earth point. The blue line in each sub-figure shows the degree of 
polarization averaged between the latitudes -5$^\circ$ and +5$^\circ$. The modulation due to ring-like patterns can be seen in the blue lines in the upper three sub-figures, 
that pertain to the 
observations that were taken first, and not in the lowest three sub-figures,
that pertain to the observations that were taken last. The strength of the ring pattern varies from filter to filter. In particular, the relatively strong visibility of the patterns in the Na filter at 589.4~nm, compared with the H--alpha and H--alpha continuum filters at longer wavelengths, is qualitatively consistent with the model prediction that the sensitivity of the polarization signal to gas-density variations increases toward shorter wavelengths because of stronger Rayleigh scattering (see Fig.~\ref{fig:polvar}). At the same time, the observations do not follow a purely wavelength-driven sequence, indicating that observing conditions, including atmospheric seeing and airmass, could have played an important role. This is not surprising because, already during the first observations (through the H-alpha filter), 
Venus had an elevation of only about 24$^\circ$ and thus low on the
hazy horizon (see Fig. \ref{fig:venusGaussianBlur} for the effect of variation in atmospheric seeing). ExPo was an experimental instrument that underwent continuous development and modification. It was never reinstalled on the telescope in an identical configuration and has since been decommissioned.

The strong feature at longitudes of about -12$^\circ$ 
in the observations through the H--alpha filter, appears to also have been 
captured in the observations through the H--alpha cont. filter,
except somewhat closer towards the terminator. The time interval between
these observations was about 10 minutes (see Table~\ref{filterspecs}).
It is tempting to derive a speed of a density wave, assuming it is indeed
the same feature, but because of the lack of continuous observations
and without any knowledge on the variability of the pattern in strength
and location, one should indeed be careful using this for the characterization
of the phenomenon.

\begin{figure}[ht]
\centering
\includegraphics[width=0.7\linewidth]{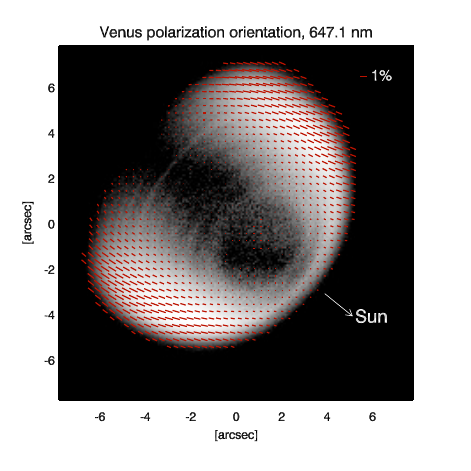}
\caption{The direction of polarization as measured using the H-alpha cont. filter
         ($\lambda_0=647.1$~nm). The orientation of the red lines indicates
         the polarization direction (note that the direction is not shown for
         every pixel to avoid cluttering), and the line length indicates the 
         local degree of polarization, which is predominantly determined
         by the polarization signal of the lower atmospheric layers that
         contain the hazes and clouds.}
\label{fig5}
\end{figure}

Figure~\ref{fig5}, finally, shows the direction of polarization as derived
from the observations through the H-alpha cont. filter. The direction is
mostly parallel to the planetary scattering plane, and thus to the equator
of the planet. The direction does not show deviations related to the rings,
which corresponds with the results of our numerical simulations 
(see Fig.~\ref{fig:polvar}), in which the polarization continuum is
negative, hence parallel to the reference plane. An increased gas density above
the clouds and hazes will force the degree of polarization closer towards
zero, but $P$ remains negative.  

The direction of polarization across the disk in Fig.~\ref{fig1} 
is mostly parallel to the planetary scattering plane, without a significant change of the direction of polarization 
across the rings (see Fig.~\ref{fig5}).
The rings appear in the images taken in the H--alpha, H--alpha continuum,
and the Sodium filters, which were taken first during the excellent seeing 
conditions. 
In the images taken in Na continuum, Sloan~r, and Sloan~i filters, the 
polarization across the poles is still high and that at low latitudes is
still low, but there are no rings distinguishable within ExPo's sensitivity.  
We have no other images of Venus with rings, as we did not have other 
opportunities to observe Venus under such advantageous seeing conditions, and, 
due to the experimental nature of ExPo, the instrument has been taken apart
and its optical components reused for other experiments. To the best of our 
knowledge, there are no other polarimeters available that combine imaging 
with ExPo's high polarimetric sensitivity.

As these rings have not been observed before, we first suspected them to be due to an instrumental effect, in particular because they appear to be roughly concentric around the brightest region on Venus' disk. However, ExPo was specifically designed to suppress instrumental polarization and other first-order systematic effects through its dual-beam exchange technique (Sect.~\ref{instrument}), in which the signal is switched between two orthogonally polarized channels and recovered through double differencing. We have not been able to identify an instrumental effect that would leave rings in the images taken in the H--alpha, H--alpha continuum, and Na filters and not in the other filters. Nor can we identify a plausible mechanism by which filter-induced polarization alone, including in the narrow-band filters, would produce the observed large-scale, spatially coherent ring pattern centered near the sub-solar point on Venus' disk. If the filters were the dominant cause of the signal, such a geometrically organized morphology would not be expected. Moreover, similar features have not appeared in any other high-precision polarization images taken with ExPo \citep{min2013color,jeffers2012direct,rodenhuis2012extreme}, nor in carefully executed optical laboratory measurements with ExPo, including tests using a white styrofoam ball to simulate Venus. We also considered optical effects in Earth's atmosphere, such as scattering by an optically thin layer of high-altitude ice particles, but failed to find an explanation for the shape and location of the pattern on Venus, for the stability of the pattern during the observations through the first three filters, and especially for the consistent direction of polarization across Venus' disk and the rings.

Although we have not identified a plausible instrumental explanation for the pattern, the present data do not allow us to exclude an unknown artefact completely. If the candidate signal is astrophysical, scattering of sunlight within Venus's atmosphere provides a plausible explanation. The lack of a similarly apparent pattern in the last three filters (see Table~\ref{filterspecs}) is more likely due to a combination of larger airmass, degraded seeing, wavelength-dependent sensitivity, and possible spectral smearing in the broader filters.

We have identified several explanations why such features, would not have been observed before. 
Firstly, while Venus has been observed with ground--based polarimeters before 
\citep{1974JAtS...31.1137H}, those observations did not have a spatial resolution
high enough to resolve any features on the planet's disk. 
Spatially resolved polarimetry of Venus was performed with the Pioneer Venus'
Orbiter Cloud Photopolarimeter (OCPP) \citep{Travis1979}. The polarimetric 
accuracy of OCPP, however, was $10^{-3}$ in the $365-935$~nm range \citep{Travis1979}, 
which would have been insufficient to detect the ring-like patterns that we have observed. 
The SPICAV instrument on the European Space Agency's 
(ESA) Venus Express (VEx) mission had polarimetric capabilities thanks 
to its opto-acoustic modulator \citep{Korablev2012,rossi2015preliminary}, 
but it covered mostly near infrared wavelengths, at which, as we will show
later, the rings would have been virtually invisible. Also, SPICAV's 
polarimetric accuracy was at most $10^{-3}$. The Japanese Space Agency's (JAXA) Akatsuki mission, which has been orbiting Venus since 2015, unfortunately does not carry a polarimeter. South Korea's twin-CubeSat CLOVE mission is planned to observe Venus with polarimetric capabilities \citep{lee2024long}. Venspec-H on ESA's EnVision mission to Venus, currently scheduled for launch in 2032, will include spectropolarimetric capability, but only at near-infrared wavelengths \citep{helbert2019venspec, neefs2024venspec}.

This observational data set has important limitations. It consists of a single serendipitous observing sequence, and no comparable ExPo observations of Venus are available. Although ExPo was designed for very high polarimetric sensitivity and to suppress first-order instrumental effects, the absolute polarization offset of these impromptu observations is only calibrated to the 1--2\% level. Moreover, ExPo was an experimental instrument that was not later reinstalled in an identical configuration and has since been dismantled. For these reasons, the present observations cannot by themselves establish a definitive detection of a new Venusian atmospheric phenomenon, and should instead be viewed as a candidate signal that requires independent confirmation.

\section{Observing and Modelling the rings in polarization}
\label{methods}

\vspace*{0.5cm}
\subsection{Definitions of fluxes and polarization}
\label{definitions}

We describe light with a Stokes (column) vector ${\bf F}$ as follows \citep{1974Hansen}
\begin{equation}
  {\bf F} = \begin{bmatrix}
                        F,
                        Q, 
                        U, 
                        V 
                     \end{bmatrix},
\label{eq_stokes}  
\end{equation}
with $F$ the total flux, $Q$ and $U$ the linearly polarized fluxes, 
and $V$ the circularly polarized flux, all in units of W m$^{-2}$.
These fluxes all depend on the wavelength $\lambda$, and we will compute them
per spectral filter. 
Regarding sunlight that is reflected by a planet, the circularly polarized flux 
$V$ is very small \citep{1974Hansen,rossi2018circpol} and it is 
not measured by ExPo.
The linearly polarized fluxes $Q$ and $U$ are defined with respect to 
a reference plane, for which we choose the planetary scattering plane, 
i.e.\ the plane containing the centers of the Sun, Venus, and the observer.

The degree of linear polarization $P$ is defined as 
\begin{equation}
   P = \frac{\sqrt{Q^2 + U^2}}{F},
\label{eq_pol}
\end{equation}
with $\sqrt{Q^2 + U^2}$ the total linearly polarized flux.
While the degree of linear polarization $P$ is independent of the  
reference plane used for Stokes parameters $Q$ and $U$, 
the direction or angle of linear polarization, $\chi$, does depend 
on that choice. Angle $\chi$ can be derived from
\begin{equation}
   \tan 2\chi = U/Q.
\end{equation}
The value of $\chi$ is chosen in the interval $[0^\circ,180^\circ \rangle$, 
and such that $\cos 2\chi$ has the same sign as $Q$ \citep{1974Hansen}. 

In case polarized flux $U$ equals zero, we can employ an alternative definition
of the degree of linear polarization that includes information about the direction
of polarization, namely
\begin{equation}
   P_{\rm s} = -Q/F,
\label{eq_pol2}
\end{equation}
If $P_{\rm s} > 0$ the direction of polarization is perpendicular to the 
reference plane and if $P_{\rm s} < 0$ it is parallel to the reference plane.

\vspace*{0.5cm}
\subsection{Expo instrument description}
\label{instrument}

\begin{figure}[ht]
\centering
\includegraphics[width=\linewidth]{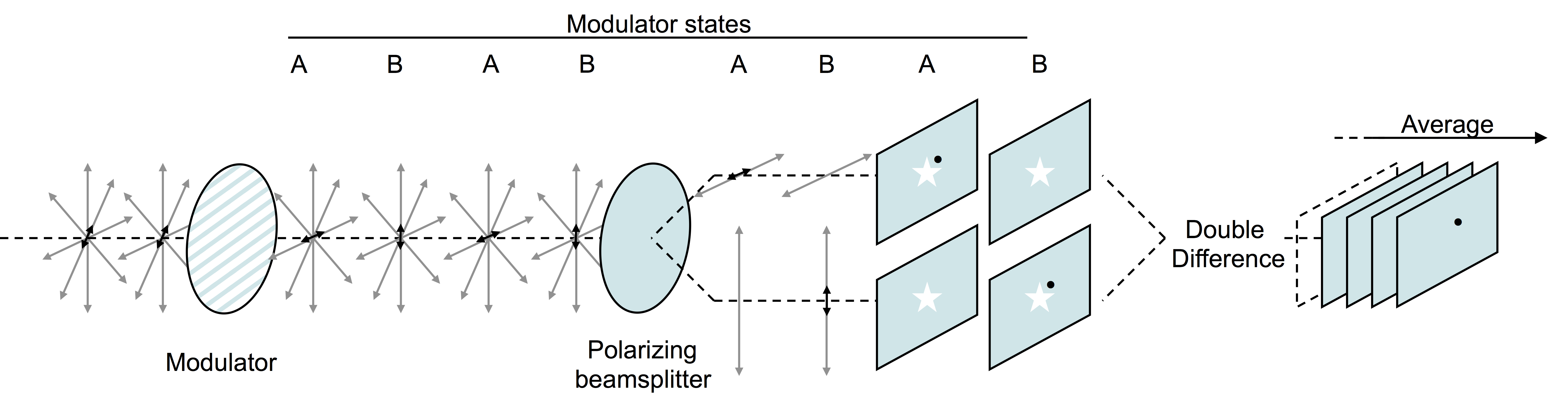}
\caption{ExPo's dual-beam exchange polarimetry technique.}
\label{fig:beamexch}
\end{figure}

Polarimetry is a differential measurement technique: in a simple polarimeter, 
a polarizing beam--splitter is used to split the light in its two orthogonally 
polarized components. Of each of these components, the flux is then measured. 
The difference of two such flux measurements yields the net linearly polarized 
flux of the incoming light along the orthogonal axes defined by the beam--splitter. ExPo \citep{rodenhuis2008extreme, canovas2011data} has been designed to avoid introducing systematic or temporal errors by using a dual-beam exchange measurement approach: the light is passed through a polarising beamsplitter and the two orthogonal polarisation states are measured simultaneously. In front of the beamsplitter, a switchable half-wave plate can interchange the two polarisation states between the two beams. In the second state, the two polarisation states are measured again. This technique is depicted in Fig.~\ref{fig:beamexch}. A Ferro-electric liquid crystal (FLC) modulator switches between the two states synchronously with the camera's 35~Hz frame rate, thus canceling first--order systematic errors due to optical path differences and reducing noise that is due to variations in atmospheric seeing between consecutive frames. This dual-beam exchange and double-differencing strategy strongly suppresses first-order instrumental polarization and makes it difficult for a static optical component, such as a filter, to imprint a coherent disk-wide polarization morphology on the final reduced images \citep{canovas2011data}.

Calling the two states $A$ and $B$ (see \ref{fig:beamexch}) and denoting the two beams by $L$ and $R$, the polarised flux is found through the double difference of four measurements:
\begin{equation}
Q = A_L-A_R-B_L+B_R
\end{equation}
Rotating the switchable half-wave plate by 22.5$^\circ$ makes the polarimeter sensitive to polarisation at 45$^\circ$ with respect to the first measurement, allowing the measurement of the $U$ component of the Stokes vector. The polarised flux is again found through the double difference:
\begin{equation}
U = [A_L-A_R-B_L+B_R]^{45^\circ}
\end{equation}
ExPo was a regular instrument on the William Herschel Telescope (WHT), part of the Isaac Newton Group of telescopes on the island of La Palma, Spain. The WHT has an experimental instrument station at one of the Nasmyth ports, where a standard optical table is available on which such instruments can be mounted.


\subsection{Radiative transfer algorithm}
\label{rtalgorithm}

Our radiative transfer computations to reproduce the rings are based on an adding--doubling algorithm
\citep{deHaan1987} that fully includes linear and circular polarization for all
orders of scattering. The algorithm requires a model atmosphere to be 
composed of a stack of horizontally homogeneous layers, and, for each of the 
layers, it requires the specification of the optical thickness $b$, the single 
scattering albedo $\omega$, and the single scattering matrix ${\bf S}$
of the scattering particles in the layer.
The atmosphere is bounded below by a horizontally homogeneous surface, 
which we assume to be black. The surface albedo is irrelevant for our simulations.

The optical thickness of an atmospheric layer at wavelength $\lambda$ \citep{stam1999degree}
is given by
\begin{equation} 
    b(\lambda) = b^{\rm m}(\lambda) + b^{\rm a}(\lambda),
\end{equation}
with $b^{\rm m}$ the layer's gaseous optical thickness and $b^{\rm a}$ the 
layer's aerosol optical thickness, which is due to the cloud or haze particles
(in our model atmosphere, a layer that contains aerosol, contains either cloud 
or haze particles, not a mixture of both). 
The single scattering albedo in the layer is given by
\begin{equation}
  \omega(\lambda) = \frac{\omega^{\rm m}(\lambda) b^{\rm m}(\lambda) +
                          \omega^{\rm a}(\lambda) b^{\rm a}(\lambda)}
                         {b^{\rm m}(\lambda) + b^{\rm a}(\lambda)},
\end{equation}
with $\omega^{\rm m}$ the single scattering albedo of the gas molecules, and 
$\omega^{\rm a}$ the single scattering albedo of the cloud or haze particles
in the layer.
The single scattering matrix of the particles in a layer is given by
\begin{equation}
   {\bf S}(\lambda) = 
   \frac{\omega^{\rm m}(\lambda) b^{\rm m}(\lambda) {\bf S}^{\rm m}(\lambda)+
         \omega^{\rm a}(\lambda) b^{\rm a}(\lambda) {\bf S}^{\rm a}(\lambda)}
        {\omega(\lambda) b(\lambda)},
\end{equation}
with ${\bf S}^{\rm m}$ the single scattering matrix of the gas and 
${\bf S}^{\rm a}$ the single scattering matrix of the cloud or haze particles.

The single scattering matrix ${\bf S}^{\rm m}$ of anisotropic gaseous 
scattering and the depolarization factor $\delta(\lambda)$ is given by \cite{1974Hansen}

\begin{equation}
{\bf S}^{\rm m}(\lambda,\Theta) = 
\Delta(\lambda)
\begin{bmatrix}
	\frac{3}{4}(1 + \cos^2 \Theta) + \frac{1}{\Delta(\lambda)} - 1 & - \frac{3}{4} \sin^2 \Theta & 0 & 0 \\
   -\frac{3}{4} \sin^2 \Theta & \frac{3}{4} (1 + \cos^2 \Theta)  & 0 & 0 \\
    0 & 0 & \frac{3}{2} \cos \Theta & 0\\
    0 & 0 & 0 & \Delta'(\lambda) \frac{3}{2} \cos \Theta
\end{bmatrix},
\label{scatmat}
\end{equation}

where

\begin{equation}
   \Delta(\lambda) = \frac{1 - \delta(\lambda)}{1 + \delta(\lambda)/2}
   \hspace*{0.5cm} {\rm and} \hspace*{0.5cm}
   \Delta'(\lambda) = \frac{1 - 2 \delta(\lambda)}{1 - \delta(\lambda)}.
\end{equation}

The four layers of our model atmosphere all contain pure carbon-dioxide, 
i.e. CO$_2$, gas.
In the filters that we used for our observations, 
CO$_2$ does not have any significant absorption bands.
We thus neglect gaseous absorption in our computations, i.e.\ 
$\omega^{\rm m}= 1.0$.
The gas optical thickness, $b^{\rm m}$, at a given 
wavelength $\lambda$, of an atmospheric layer is then computed according to
\begin{equation}
   b^{\rm m}(\lambda) = N^{\rm m} \sigma^{\rm m}(\lambda),
\end{equation}
with $N^{\rm m}$ the gas column number density (in m$^{-2}$) in the layer,
and $\sigma^{\rm m}$ the gaseous extinction cross--section (in m$^2$),
which in our case equals the gaseous scattering cross--section, due to the 
lack of absorption. This cross--section is given by
\begin{equation}
  \sigma^{\rm m}(\lambda) = 
       \frac{24 \pi^3}{N_{\rm L}^2} \frac{(n^2(\lambda)-1)^2}{(n^2(\lambda)+2)^2}
       \frac{(6 + 3\delta(\lambda))}{(6 - 7 \delta(\lambda))} 
       \frac{1}{\lambda^4} ,
\end{equation}
with $N_{\rm L}$ Loschmidt's number, $n$ the refractive index of CO$_2$ under
standard conditions, and $\delta$ the depolarization factor of pure CO$_2$.
For $n$ and $\delta$ of pure CO$_2$ gas, we assume the wavelength dependent 
relations presented by  
\cite{2005JQSRT..92..293S}. It should be noted that the error in the equation for $n$ by \cite{2005JQSRT..92..293S} for CO$_2$ is  $\sim$4~$\%$.

Assuming that each atmospheric layer is in hydrostatic equilibrium, a layer's
gas column number density $N^{\rm m}$ (in m$^{-2}$) is computed using
\begin{equation}
   N^{\rm m} = N_{\rm A} \frac{p_{\rm bot}-p_{\rm top}}{m g},
\end{equation}
with $N_{\rm A}$ Avogadro's number, 
$p_{\rm bot}$ and $p_{\rm top}$ the pressures at the bottom and at the 
top of the atmospheric layer, respectively,
$m$ the average molar mass of the gas, for which we use the value for pure CO$_2$, 
i.e. 44.01 g.mol$^{-1}$, $g$ the acceleration of gravity, which we assume to be 
altitude independent across the region of Venus' atmosphere that we study
and equal to 8.87~m.s$^{-2}$.
Table~\ref{tab2} shows the pressures at the bottom and the top of each 
of the four atmospheric layers that we use in our model, the 
computed gas column number density 
$N^{\rm m}$ and gaseous optical thickness $b^{\rm m}$ 
at 580~nm (the central wavelength of the Na cont. filter) for each of the 
atmospheric layers. We impose the variations in the gas column number density on the highest 
atmospheric layer. Thus, we impose a perturbation $\delta N^{\rm m}$ on the top layer, with a baseline gas column number density of $N^{\rm m} = 6.942 \times 10^{27}\,\mathrm{m}^{-2}$ (see Table~\ref{tab2}), corresponding to the atmospheric column above the cloud tops.

\begin{table}
\caption{The depolarization factor $\delta$ and the 
         the refractive index $n$ of CO$_2$ \citep{2005JQSRT..92..293S}, 
         and the molecular
         scattering optical thickness $b^{\rm m}_{\rm tot}$ of the standard 
         model atmosphere as a whole at the central wavelength $\lambda_0$ 
         of each filter (cf.\ Table~\ref{filterspecs}).
         Of the $b^{\rm m}_{\rm tot}$ of the atmosphere, 
         98.92\% pertains to layer 1 (the bottom layer), 1.022\% to layer 2,
         0.000484\% to layer 3, and 0.0000538\% to layer 4 (the top layer).
         Furthermore, $\sigma^{\rm a}_{\rm cloud}$ is the scattering cross--section 
         of the cloud particles, $b^{\rm a}_{\rm cloud}$ the optical thickness
         of the cloud, $\sigma^{\rm a}_{\rm haze}$ the scattering cross-section of 
         the haze particles, and $b^{\rm a}_{\rm haze}$ the optical thickness
         of the haze.
         }
\centering
\resizebox{0.95\textwidth}{!}{
\begin{tabular}{c c c c c c c c c}
\hline 
Filter & $\lambda_0$ {\small [nm]} & $\delta$ & $n$ & $b^{\rm m}_{\rm tot}$ & 
	$\sigma^{\rm a}_{\rm cloud}$ {\small [$\mu$m$^2$]} & $b^{\rm a}_{\rm cloud}$ &
    $\sigma^{\rm a}_{\rm haze}$ {\small [$\mu$m$^2$]} & 
    $b^{\rm a}_{\rm haze}$ \vspace*{0.05cm} \\
\hline
Na cont.      & 580.0 & 0.0784 & 1.0004262 & 13.43 & 6.635 & 30.00 & 0.154 & 0.020 \\
Na            & 589.4 & 0.0783 & 1.0004260 & 12.57 & 6.634 & 29.99 & 0.151 & 0.020 \\
Sloan r       & 623.1 & 0.0779 & 1.0004251 &  9.41 & 6.638 & 30.01 & 0.141 & 0.018 \\
H-alpha cont. & 647.1 & 0.0772 & 1.0004246 &  8.59 & 6.650 & 30.06 & 0.134 & 0.017 \\
H-alpha       & 656.3 & 0.0776 & 1.0004244 &  8.11 & 6.657 & 30.10 & 0.131 & 0.017 \\
Sloan i       & 762.5 & 0.0769 & 1.0004228 &  4.41 & 6.928 & 31.32 & 0.104 & 0.014 \\
\hline    
\end{tabular}}
\label{tab0}    
\end{table}

\begin{table}[hb]
\centering
\caption{The parameters of our standard Venus model atmosphere, addressing the 
         layers from top to bottom: $p_{\rm bot}$ and $p_{\rm top}$ are the 
         standard pressures at the bottom and the top of each layer, $N^{\rm m}$ 
         is the standard gas column number density based on the pressure range
         across each layer, $b^{\rm m}$ is the gaseous  
         optical thickness of the layer at 580~nm, the central wavelength of the
         Na cont. filter (see Table~\ref{filterspecs}), and $b^{\rm a}$ is the
         cloud or haze optical thickness of the layer, also at 580~nm.
         Layer~2 contains the cloud and layer~3 contains the haze
         (see Table~\ref{tab0} for the values at other wavelengths).
         The single scattering albedo's of
         the gaseous molecules, $\omega^{\rm m}$, and that of the cloud and haze
         particles equals 1.0 in each layer and at each wavelength.}
\resizebox{0.7\textwidth}{!}{
\begin{tabular}{c c c c c c}
\hline 
Layer & $p_{\rm bot}$ (bars) & $p_{\rm top}$ (bars) & $N^{\rm m}$ (m$^{-2}$) & $b^{\rm m}(580~{\rm nm}) $ & $b^{\rm a}(580~{\rm nm})$ \\
\hline
4 & 0.005 & 0.0 & 7.713e26 & 0.000794 & 0.0 \\
3 & 0.05 & 0.005 & 6.942e27 & 0.00650 & 0.02 \\
2 & 1.0 & 0.05  & 1.466e29 & 0.137 & 30.0 \\
1 & 93.0 & 1.0 & 1.419e31 & 13.283 & 0.0 \\
\hline    
\end{tabular}}
\label{tab2}    
\end{table}

The second and third layers of the model atmosphere contain, respectively, cloud
and haze particles (see Table~\ref{tab2}). These particles are assumed to be 
spherical and homogeneous. For both
particle types, we assume a two parameter gamma size distribution \citep{1974Hansen}.
The cloud particles have an effective radius $r_{\rm eff}$ of $1.05~\mu$m and 
an effective variance $\nu_{\rm eff}$ of 0.07 \citep{1974JAtS...31.1137H},
and the haze particles, $r_{\rm eff}=0.25~\mu$m and $\nu_{\rm eff}= 0.25$ \citep{Sato1996}. 
We assume that the cloud and haze particles consist of 75\% sulfuric 
acid \citep{1974JAtS...31.1137H}, and adopt the following wavelength dependence for their refractive index 
$n_{\rm r}$
\begin{equation}
   \frac{n_{\rm r}^2(\lambda) - 1}{n_{\rm r}^2(\lambda) + 2} = 
        \frac{-29.8034}{\lambda^{-2} - 116.3899}
\end{equation}
as derived from the values provided
by Hansen \& Hovenier \citep{1974JAtS...31.1137H}.

For each central filter wavelength $\lambda_0$ (Table~\ref{filterspecs}), 
we compute the optical properties of the cloud and haze particles, 
i.e.\ their extinction cross--section $\sigma^{\rm a}$, their single scattering matrix ${\bf S}^{\rm a}$,
using a Mie--algorithm \citep{deRooij1984}. 
Their single scattering albedo $\omega^{\rm a}$ equals 1.0 due to the lack of
absorption.

\begin{figure}[!h]
\includegraphics[width=0.5\textwidth]{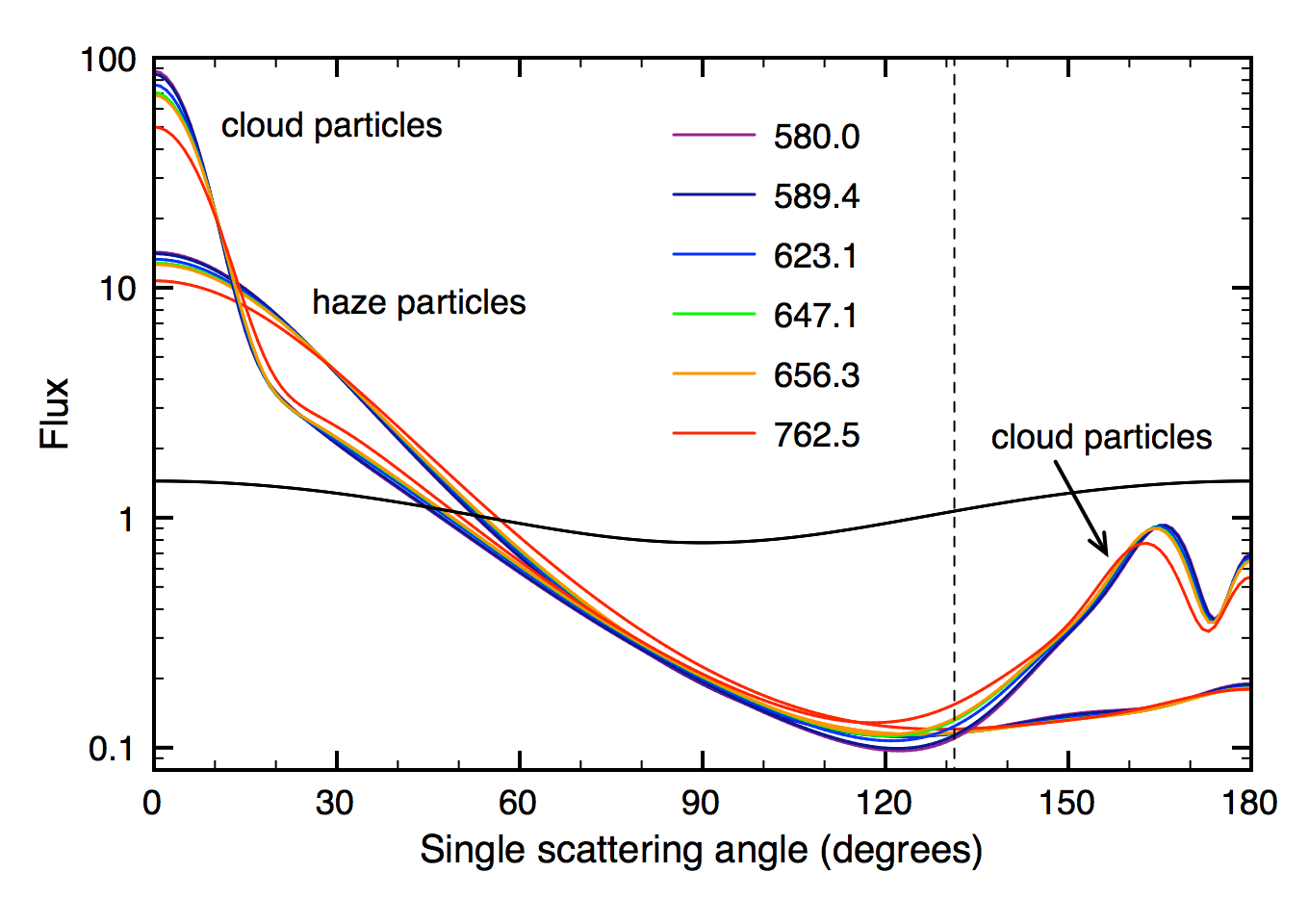}
\includegraphics[width=0.5\textwidth]{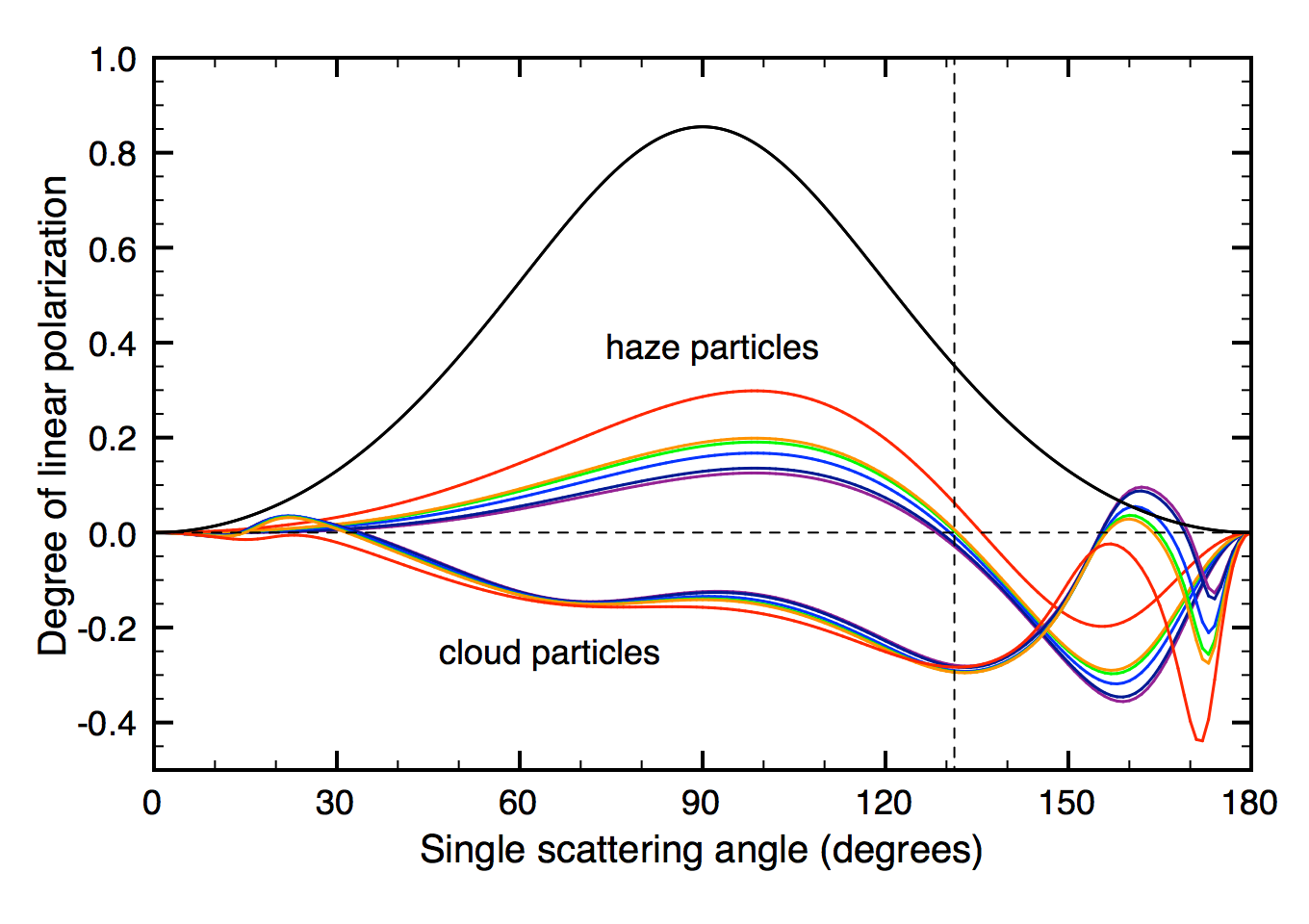}
\caption{The flux (left) and degree of linear polarization $P$ (right) of 
         incident unpolarized light that is singly
         scattered by the cloud particles, haze particles, and
         gaseous molecules (solid black line) at the central wavelengths of
         the filters. Note that the wavelength dependence of the molecular
         scattering (through the depolarization factor, see Eq.~\ref{scatmat}) is
         so small that we show only one line. All flux curves are normalized
         such that averaged over all scattering directions equals 1.0. 
         A positive $P_s$ indicates a direction of
         polarization perpendicular to the scattering plane (the plane
         containing the incident and scattered light beams) and a negative
         $P_s$, a direction parallel to the scattering plane. The black, dashed
         vertical lines indicate the single scattering angle that corresponds
         with the phase angle of Venus at the time of our observations,
         and the black, dashed horizontal line in the polarization plot, $P=0$.}
\label{fig_single}
\end{figure}

Figure~\ref{fig_single} shows the total flux and degree of linear
polarization of unpolarized incident light that
is singly scattered by the cloud and haze particles for the central filter
wavelengths $\lambda_0$. For comparison,
we have also included the curves for the light that is scattered by
the gaseous molecules. As can be seen, at a single scattering angle of 131.3$^\circ$, which corresponds to Venus' phase angle of 48.7$^\circ$ at the time of the ExPo observations, the light scattered by the cloud particles is polarized parallel to the reference plane (which would correspond to the planetary scattering plane, which also contains the planet's equator), while the light scattered by the gaseous molecules is polarized perpendicular to this plane. For the haze particles, the degree of polarization at this scattering angle is close to zero for most wavelengths shown in Fig.~6, and their polarization direction is therefore effectively undefined; at 762.5~nm, however, it is weakly polarized perpendicular to the reference plane.

We choose the cloud and haze optical thickness, $b^{\rm a}$, of 
atmospheric layers two and three according to previous Venus atmospheric 
models \citep{kawabata1980cloud,rossi2015preliminary}.
Table~\ref{tab2} lists the optical thickness $b^{\rm a}$ of each layer 
at 580~nm (which is the central wavelength $\lambda_0$ of the Na cont. filter). 
The cloud or haze optical thicknesses at other wavelengths are computed 
according to
\begin{equation}
   b^{\rm a}(\lambda) = b^{\rm a}(580~{\rm nm})
                        \frac{\sigma^{\rm a}(\lambda)} 
                             {\sigma^{\rm a}(580~{\rm nm})}.
\end{equation}
The values of $b^{\rm a}$ of the cloud ($b^{\rm a}_{\rm cloud}$) and 
the haze ($b^{\rm a}_{\rm haze}$) at the central filter
wavelengths can be found in Table~\ref{tab0}. The particle scattering properties at the other wavelengths are calculated using the Mie-code as described earlier. 

We assume that the sunlight that is incident on the planet is 
unpolarized \citep{1987kemp}, and it is thus described by the column 
vector $F_0 [1,0,0,0]$, with $F_0$ equal to $\pi$.

To numerically simulate the observations, 
we represent Venus' disk with a grid of equally sized, square pixels. We projected the planetary disk onto the plane of the sky and divided the circumscribing square into a 300 × 300 grid, with the equator aligned to the horizontal x-axis. The center of each pixel was mapped onto the spherical planet to determine the location for computing the locally reflected Stokes vector. For each
pixel, we compute the Stokes vector ${\bf F}$ (see Eq.~\ref{eq_stokes}) of the 
locally reflected sunlight with wavelength $\lambda$, according to 
\begin{equation}
    {\bf F}(\lambda,\theta,\theta_0,\phi-\phi_0) = 
        \cos \theta_0 \hspace*{0.1cm}  
        {\bf R}(\lambda,\theta,\theta_0,\phi-\phi_0) \hspace*{0.1cm}
        {\bf F}_0(\lambda),
\label{eqrefF}        
\end{equation}
where $\theta_0$ is the local solar zenith angle, $\theta$ the local zenith 
viewing angle, and $\phi - \phi_0$ the local azimuthal difference angle
\citep{deHaan1987}.
Vector ${\bf F}_0$ is the incident sunlight, and ${\bf R}$ is
the local planetary reflection matrix. Matrix ${\bf R}$ depends on the composition 
and structure of the local atmosphere and surface, and given these properties
as described above, we compute it using an efficient adding--doubling 
algorithm \citep{deHaan1987} that fully includes linear 
and circular polarization for all orders of scattering.

All angles are computed by projecting the center of each pixel onto a 3D--sphere
that represents Venus. Note that while $\theta$ and $\phi$ depend on the position
of the projected center of each pixel with respect to the sub--observer point 
on the sphere,
$\theta_0$ and $\phi_0$ depend on the position of the projected center of each
pixel with respect to the sub--solar point, the location of which depends on
Venus' phase angle $\alpha$, which equaled 48.7$^\circ$ at the time of our 
observations.

The reflected fluxes as computed using Eq.~\ref{eqrefF} do not need
further scaling because the pixels all have the same size.
The lack of absolute calibration of our observed total and polarized 
fluxes, precludes an attempt to absolutely fit the computed fluxes.
The degree of polarization (Eq.~\ref{eq_pol}) is a relative measure 
and as such it is in principle insensitive to e.g.\ changes in the airmass during 
the observational cycle. While the measurements by ExPo have a very high 
polarimetric sensitivity reaching 10$^{-4}$ \citep{rodenhuis2012extreme}, 
the absolute calibration of the degree of 
polarization of our observations, however, is only 1 - 2\%. 
In other words, the absolute off--set of the degree of polarization is 
on the order of 1 - 2\% ($P_{\rm observed} = P_{\rm Venus} \pm 1 - 2$\%). 
As we will show below, this off--set 
does not limit our numerical simulations of the observations, as those
are based on the spatial variability of $P$ and not on its absolute value.

\section{INTERPRETING THE RING-LIKE POLARIZATION PATTERN}
\label{observation_explanation}

The degree and direction of polarization of sunlight that is reflected by a 
region on a planet depends strongly on the optical properties of the 
scattering particles, on the illumination and viewing geometries, and on the 
ratio of singly to multiply scattered light \citep{1974Hansen}, as the former 
can be highly polarized while the latter usually has a low degree of polarization. 
Because of the large distances between Venus and the Sun and between 
Venus and Earth, and given the phase angle of 48.7$^\circ$,
the singly scattered photons that ExPo received from locations 
across Venus' disk have been scattered across an angle of $131.3^\circ$.
If the rings were due to variations in the microphysical properties
of the scattering haze and/or cloud particles in the atmosphere, such as their 
size or composition, these variations should appear in a similar, concentric 
planet--wide pattern and they should be small enough to virtually leave no 
trace in the total flux.
This seems difficult to explain with our knowledge of the 
composition and structure of the clouds and hazes in Venus' atmosphere.
The rings could also be due to variations in the local number densities of haze
and/or cloud particles, as such variations would change the ratio of singly to
multiple scattered light and thus
the degree of 
polarization of
the reflected
light. However, the known, latitudinal variations
in the cloud and haze properties (notably, cloud top altitude \citep{Ignatiev2009} 
and haze optical thickness \citep{kawabata1980cloud,Sato1996,rossi2015preliminary})
seem to preclude the appearance of the concentric, planet--wide patterns.

The regularity of the polarization pattern and its apparent superposition on the high-polarization polar regions are consistent with, though do not prove, an origin in Venus's upper atmosphere. If the observed pattern is astrophysical rather than instrumental, one plausible explanation is spatial variation in the gas density above the hazes and clouds. Such planet--wide density variations could result 
from atmospheric waves emanating from the region close to the sub--solar point, 
as we will discuss later.
Small, local density variations in the gas above the hazes and clouds 
will yield small variations in the amount of locally reflected sunlight, 
particularly in the amount of light that has been singly scattered by
gas molecules, which has a relatively high degree of polarization at a 
scattering angle of 131.3$^\circ$ (see\ Fig.~\ref{fig_single} for a detailed explanation), with a direction perpendicular to the scattering
plane, i.e.\ opposite to that of the clouds.  

\begin{figure}[t]
\centering
\includegraphics[trim=0.2cm 0cm 0.5cm 1cm,clip=True,width=8.7cm]{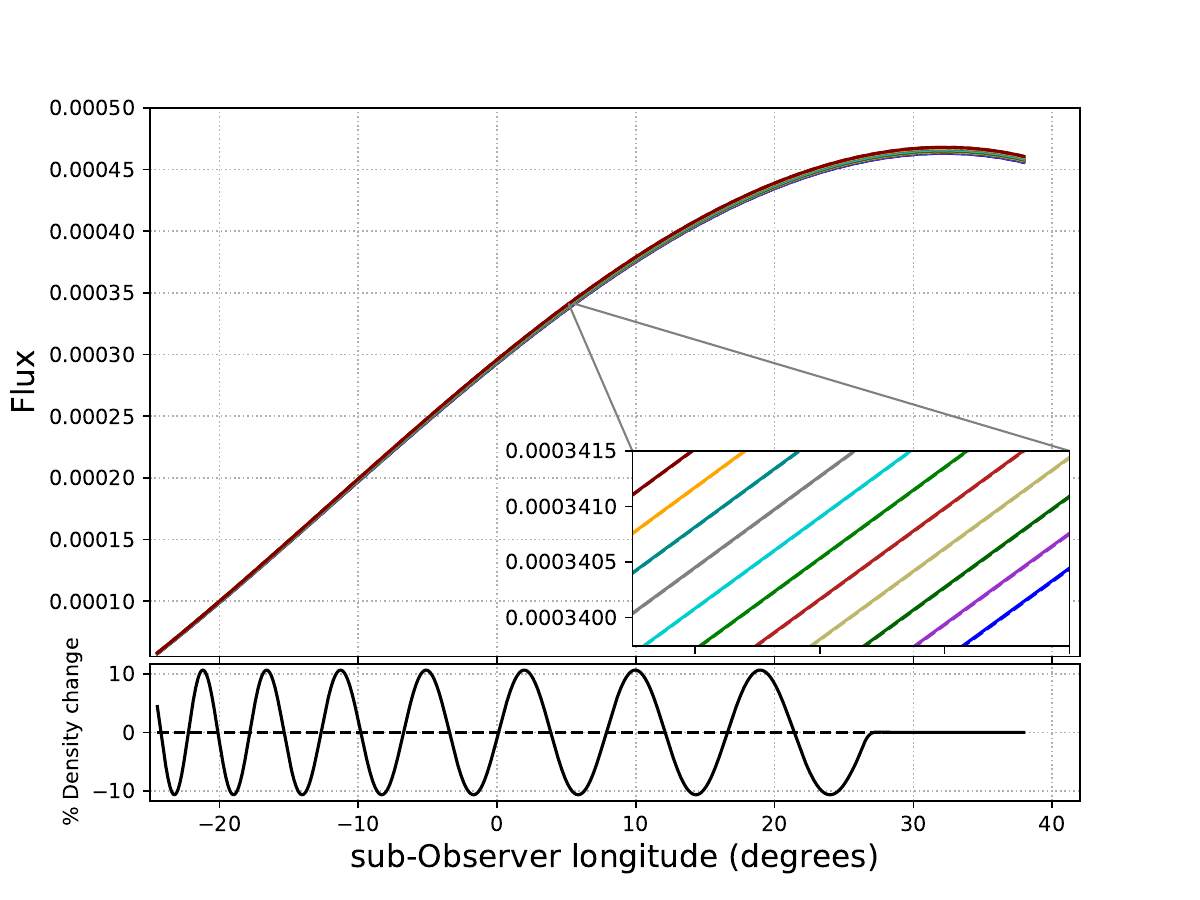}
\includegraphics[trim=0.2cm 0cm 0.5cm 1cm,clip=True,width=8.7cm]{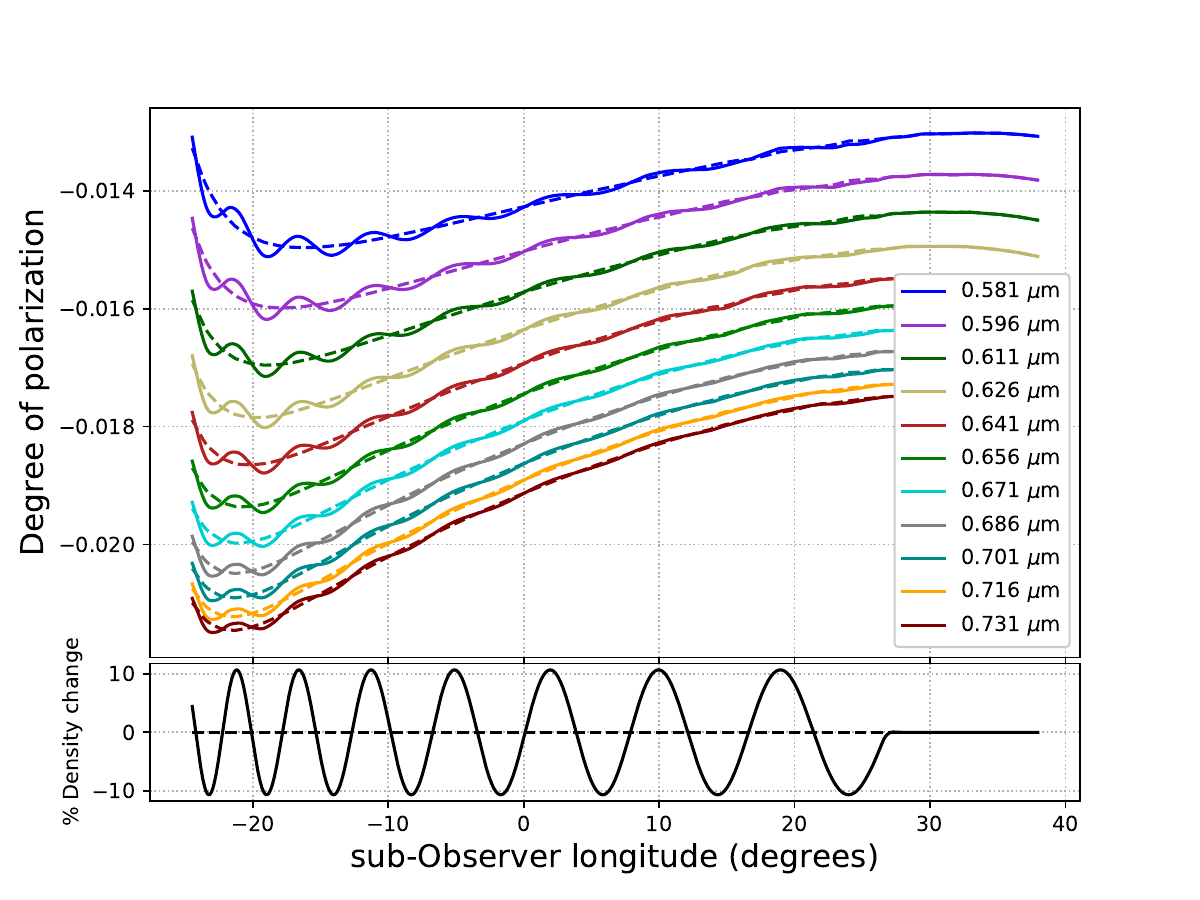}
\vspace*{0.2cm}
\caption{The total flux (left) (normalized such that at a phase angle of 0$^\circ$,
         the disk--integrated flux equals the planet's geometric albedo)
         and degree of linear polarization (right) along Venus' equator 
         in response to changes in the gas column number density 
         $N^{\rm m}$ of the upper atmospheric layer for wavelengths 
         covered by the filters (see Table~\ref{filterspecs}).
         The composition and structure of the lower atmospheric 
         layers (with cloud and haze particles) are the same at every 
         longitude. 
         The dashed lines pertain to the baseline value of 
         $N^{\rm m}$, i.e. $7.0~\times~10^{27}$ molecules m$^{-2}$
         (for the atmosphere up from 70~km or down to 0.036 bars). 
         The negative sign of the degree of polarization indicates
         a direction parallel to the planetary scattering plane 
         (and the equator). The empirical model shows that wave wavelengths vary from about 100 km near the terminator, gradually increasing to around 900 km near the sub-solar point. 
         }
\label{fig:polvar}
\end{figure}

\begin{figure}[ht]
\centering
\includegraphics[width=0.99\linewidth]{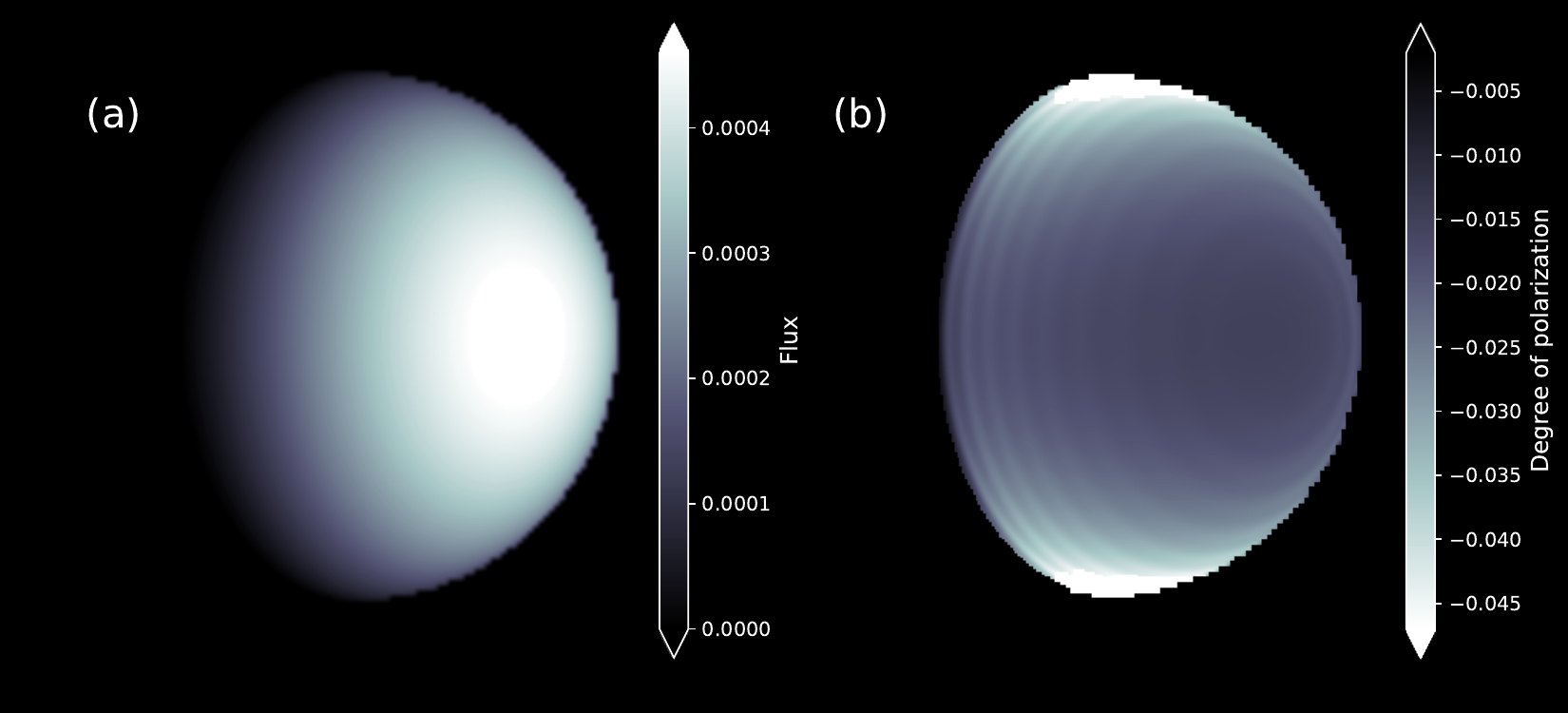}
\caption{Numerical computations corresponding to the candidate pattern in the H--alpha filter
         ($\lambda_0=656$~nm), for the total flux (left) and the polarized flux
         (right). The background polarization signal has been subtracted from the 
         latter. The decrease in spatial resolution due to the Earth's 
         atmosphere is not included. 
         The planetary atmosphere is assumed to be horizontally 
         homogeneous except for the gas density variations.}
\label{fig:venuscomputed}
\end{figure}

\subsection{Numerically reproducing the rings}

Figure~\ref{fig:polvar} shows numerical simulations of the flux and degree of 
polarization along Venus' equator at a planetary phase angle of 48.7$^\circ$. The purpose of the radiative-transfer calculations is not to demonstrate that the observed pattern must be atmospheric in origin, but to assess whether a physically plausible atmospheric perturbation could produce a signal of the observed type.

For the computations, we varied $N^{\rm m}$, the gas column number density 
(in m$^{-2}$) in the highest atmospheric layer.
The dashed lines in Fig.~\ref{fig:polvar} represent the flux and polarization
of the reflected sunlight without gaseous density variations. 
A negative degree of polarization $P_s$
indicates a polarization direction parallel to the scattering plane. 
The reflected flux appears to be virtually
insensitive to variations in $N^{\rm m}$, just like we observed. However, increasing $N^{\rm m}$ adds positive polarized flux (because of Rayleigh scattering), thus $P_s$ decreases, in the sense that $P_s$ gets closer to zero. If the background $P_s$ would be positive, increasing $N^{\rm m}$ would increase $P_s$. Indeed, the computed variations in $P_s$ are within ExPo's sensitivity and, even though
the direction of polarization of the light scattered by the molecules is
perpendicular to the reference plane and thus counteracts the polarization
direction of the clouds and hazes, the
direction of polarization across the variations remains parallel to 
the reference plane, like in the observations. 

The variations in $P_s$ are independent of the precise vertical location of 
the density variations,
except that they should be in the upper region of the atmosphere,
where the scattering is dominated by the gas.
The decrease of the scattering--cross section of the gas molecules
with wavelength (see \cite{1974Hansen} and Sec. \ref{methods})
causes the decrease of the sensitivity of $P_s$ to the density variations. This sensitivity decrease could contribute to the lack of an obvious ring-like pattern in the 
longer wavelength filters, such as the Sloan i filter, in addition to 
the larger airmass and worse seeing conditions during those observations
(see Table~\ref{filterspecs}). We did not optimize the clouds and hazes in our model atmosphere
to fit the background flux and polarization in the ExPo observations,
because, as mentioned earlier, the absolute accuracy of the 
polarization observations is limited to 1-2~\%. 
This, however, poses no problem for our interpretation, because 
the absolute change in $P_s$ due to a
small change in $N^{\rm m}$ appears to be insensitive to the 
background polarized flux as it adds mostly singly scattered light,
which does change the reflected polarized flux, but hardly
changes the total flux.

Figure~\ref{fig:venuscomputed} shows numerical simulations of the total 
flux, degree and direction of polarization across Venus' disk in the 
H-alpha filter with the density variations chosen to be concentric  
around the region close to the sub--solar point. For these simulations,
the planet was assumed to be horizontally homogeneous except for the
gas column number density in the upper atmospheric layer (the cloud
and haze macro and micro-physical properties are thus the same everywhere).

A sinusoidal variation in the gas column number density was spatially introduced to model the effect of a density wave travelling through the upper atmosphere of Venus (see lower sub-plots of Fig. \ref{fig:polvar}). The amplitude of this variation was 10 $\%$ of the computed column gas density pertaining to this top most atmospheric layer. This spatial plot illustrates that a 10\% variation in the gas density of the upper atmospheric layer can produce a significant variation in $P$ while leaving $F$ nearly unchanged. The model wavelengths (for the rings) were chosen to span the full range of gravity wave scales observed in Venus’ atmosphere, varying from about 900 km near the sub-solar (sub-observer) point to ~100 km near the terminator. Although empirical in nature, this choice is consistent with observations from ExPo and the Pioneer Venus mission, as well as with theoretical predictions (e.g., \cite{kasprzak1988wavelike, alexander1992mechanism}). More recently, the in-situ Venus Express Aerodynamic Drag Experiment revealed gravity wave activity in the planets thermosphere ($\sim$130-140 km) with wavelengths varying between 100 and 300 km \citep{muller2016situ}.

\section{Discussion and conclusion}
\label{results_discussion}

We have examined a candidate ring-like polarization pattern in a unique serendipitous ExPo data set of Venus and explored whether such a pattern could plausibly arise from gas-density variations above the cloud tops. Similar to the observations, the modeled density variations do not leave a significant signal in the reflected total flux. If the candidate pattern is astrophysical, density perturbations of order $\pm$10\% above the cloud and haze layers (i.e.\ above 0.036 bars) could account for the observed polarization modulation. One possible interpretation is that such perturbations are related to previously unrecognized planet-wide atmospheric waves excited near the cloud tops about 20$^\circ$ (~2100~km) downwind from the sub-solar point (corresponding to a 
local time of roughly 13:00~h).

This displacement with respect to the sub-solar point
could be the result of a combination of a time lag in the maximum heating of 
the cloud layer \citep{belton1976cloud} and the large equatorial wind velocities, on the order 
of 100~m/s \citep{counselman1980zonal,limaye1981cloud,newman1984zonal,moissl2009venus}
that move the clouds away from the sub-solar point.
The waves could be propagating across the planet in the direction of the 
anti-solar point, possibly within the known sub-solar to anti-solar 
flow \citep{goldstein1991absolute,bougher1997upper}.
According to the spatial resolution of the observations, the horizontal wavelength 
of the waves would be between 100 km and 1000 km, decreasing with distance 
from the excitation region.

In-situ measurements taken over ~600 orbits,
between about 160 and 200~km altitude by the ONMS instrument on Pioneer 
Venus orbiter revealed waves (variations in gas density) in various regions on the 
planet \citep{kasprzak1988wavelike}. These waves, with wavelengths of 100--600 km
were tentatively identified as gravity waves \citep{mayr1988gravity}. 
Interestingly, the Pioneer Venus measurements showed little wave activity towards the 
sub-solar region and increasing wave activity towards the morning and 
evening terminators, which agrees well with our ground-based observations. 
A planet-wide propagation was suspected for these waves \citep{mayr1988gravity}, 
although this could not be confirmed from Pioneer Venus' 
along-track data available at the time.

Because this is our only set of Venus observations with ExPo, we do not know whether the candidate pattern reflects a regularly occurring atmospheric phenomenon. If it is indeed related to the upper-atmospheric density waves observed by Pioneer Venus, it might represent a recurrent feature of Venus's atmosphere. More observations, with ground-based telescopes and/or from a future orbiter 
outfitted with highly sensitive polarimeters, should be performed to get 
insight into their appearance and variations therein. High-spectral resolution polarimetry will be done by Venspec-H on ESA’s EnVision mission to Venus (to be launched in 2032). However, this instrument will operate in the near-infrared where the expected signals from such upper-atmospheric waves are weak.

\begin{figure}[!ht]
\centering
\includegraphics[width=0.99\linewidth]{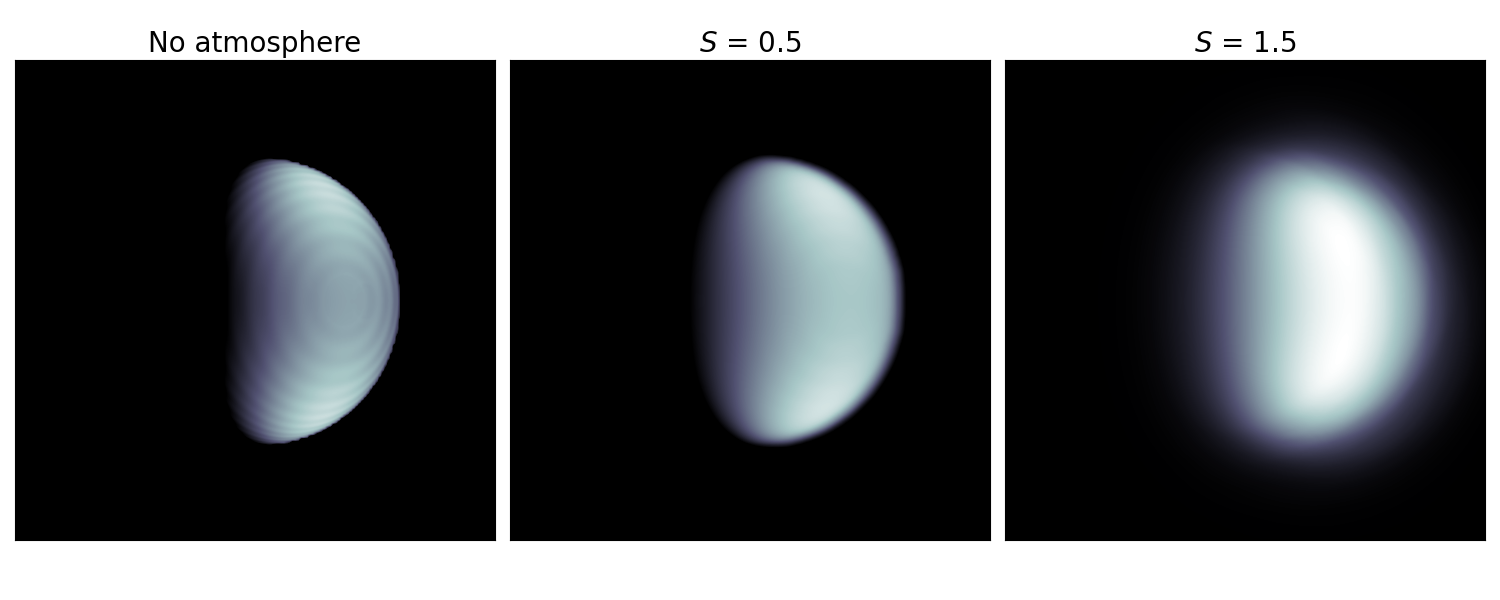}
\caption{Numerical computations of the polarized flux $Q$ in our Venus model in the H--alpha filter ($\lambda_0 = 656$ nm) after the introduction of atmospheric seeing $S$ (arcseconds). An atmospheric phase screen is applied to the fluxes in orthogonally polarized states, which are then recombined to obtain $F$ and $Q$; the signed degree of linear polarization is computed as $P_s=-Q/F$ (Eq.~4). Panels 2 and 3 show the progressive blurring of the modeled ring-like pattern for $S=0.5$ arcseconds and $S=1.5$ arcseconds, respectively.}
\label{fig:venusGaussianBlur}
\end{figure}

\begin{figure}[!ht]
\centering
\includegraphics[width=0.7\linewidth]{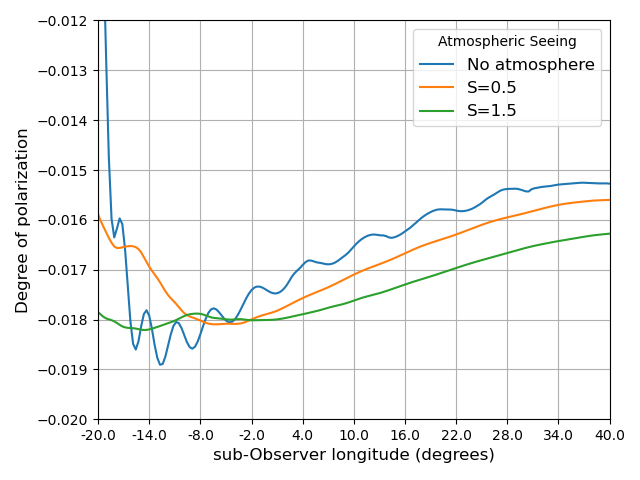}
\caption{Illustrative 1D numerical computations of the effect of Earth atmospheric seeing $S$ (arcseconds) on the modeled signed degree of linear polarization $P_s$ in the H--alpha filter ($\lambda_0 = 656$ nm). The atmosphere is assumed to be horizontally homogeneous except for the imposed gas-density variations. Increasing $S$ progressively attenuates the wave-like modulation in $P_s$. These calculations are intended to illustrate the sensitivity of the signal to seeing, not to reproduce the exact atmospheric conditions during the ExPo observations.}
\label{fig:venusGaussianBlur1d}
\end{figure}

Our numerical simulations (Figs.~\ref{fig:venusGaussianBlur} and~\ref{fig:venusGaussianBlur1d}) indicate that atmospheric blurring can substantially attenuate the polarization signal caused by the imposed gas-density variations. To explore this effect, we used an atmospheric phase screen to simulate the propagation of the incoming orthogonally polarized fluxes through Earth’s atmosphere, following the approach described by \citet{buscher2016simulating} and the implementation of \citet{mortimer2013}. The phase screen is parameterized by the atmospheric seeing $S$ (in arcseconds, specified at 500~nm), which controls the extent of image degradation caused by atmospheric turbulence and viewing geometry. For each value of $S$, the orthogonally polarized fluxes were propagated through a stochastic phase screen and then recombined to obtain the Stokes fluxes $F$ and $Q$, from which we computed the signed degree of linear polarization $P_s=-Q/F$. The simulations shown here are intended to illustrate the sensitivity of the modeled ring signal to atmospheric seeing, rather than to reconstruct the exact turbulence conditions during the ExPo observations. Because the phase screens are stochastic and the detailed atmospheric conditions during the observing sequence are not known, Figs.~\ref{fig:venusGaussianBlur} and~\ref{fig:venusGaussianBlur1d} should not be interpreted as defining a sharp detect/non-detect threshold. We therefore conclude only that increased airmass and degraded seeing could have contributed significantly to suppressing the visibility of the rings in the later observations.

The absence of the rings in the later Na continuum, Sloan $r$, and Sloan $i$ observations is likely due to a combination of effects rather than seeing alone. In particular, the sensitivity of the polarization signal to gas-density variations decreases with wavelength because of the wavelength dependence of Rayleigh scattering, and the broad Sloan filters can further reduce the observable modulation by averaging the signal over a wider wavelength range. In addition, ExPo did not have an operational atmospheric dispersion corrector during these observations, which may have introduced spectral smearing in the broadband data. The fact that the Na image at 589.4~nm still shows clear rings, despite being taken later in the observing sequence and at higher airmass than the H$\alpha$ and H$\alpha$ continuum observations, further indicates that airmass alone cannot account for the detect versus non-detect behavior, and is qualitatively consistent with the stronger modeled response at shorter wavelengths.

Our simulations also suggest that the degree of polarization due to the rings should be strongest at 
phase angles around 90$^\circ$ (our observations pertain to 48.7$^\circ$), 
because the difference between the degree of polarization of light 
singly scattered by gaseous molecules and the background polarization
is largest.
To confirm the ring--like shape of the pattern, observations at either
smaller phase angles than ours (where the strength of the pattern would, however,
be smaller), or of Venus' morning hemisphere (our observations mostly cover the
afternoon-evening hemisphere) are required. More observations, combined with an appropriate numerical model of Venus' tenuous upper atmosphere, could yield valuable insights into the physical characteristics of the waves, their surrounding environment, and their role in the planet's atmospheric dynamics.


\begin{acknowledgments}
G.M. was supported by the User Support Programme Space Research (project ALW-GO 15-37) of the Netherlands Organisation for Scientific Research (NWO). L.R. received funding from the Planetary and Exoplanetary Science (PEPSci) Programme of NWO. M.R. acknowledges support from NWO for the Extreme Polarimeter (ExPo). We thank the anonymous reviewers for their constructive comments and suggestions, which helped improve the manuscript.
\end{acknowledgments}

\newpage
\bibliography{bibliography}{}

\begin{thebibliography}{}
\expandafter\ifx\csname natexlab\endcsname\relax\def\natexlab#1{#1}\fi
\providecommand{\url}[1]{\href{#1}{#1}}
\providecommand{\dodoi}[1]{doi:~\href{http://doi.org/#1}{\nolinkurl{#1}}}
\providecommand{\doeprint}[1]{\href{http://ascl.net/#1}{\nolinkurl{http://ascl.net/#1}}}
\providecommand{\doarXiv}[1]{\href{https://arxiv.org/abs/#1}{\nolinkurl{https://arxiv.org/abs/#1}}}

\bibitem[{M. Alexander(1992)Alexander}]{alexander1992mechanism}
Alexander, M. 1992, \bibinfo{title}{A mechanism for the Venus thermospheric
  superrotation,} Geophysical research letters, 19, 2207

\bibitem[{M. Alexander {et~al.}(1993)Alexander, Stewart, Solomon, \&
  Boucher}]{alexander1993local}
Alexander, M., Stewart, A., Solomon, S., \& Boucher, S. 1993,
  \bibinfo{title}{Local time asymmetries in the Venus thermosphere,} Journal of
  Geophysical Research: Planets, 98, 10849

\bibitem[{M.~J. Belton {et~al.}(1976)Belton, Smith, Schubert, \&
  Del~Genio}]{belton1976cloud}
Belton, M.~J., Smith, G.~R., Schubert, G., \& Del~Genio, A.~D. 1976,
  \bibinfo{title}{Cloud patterns, waves and convection in the Venus
  atmosphere,} Journal of the Atmospheric Sciences, 33, 1394

\bibitem[{S. Bougher {et~al.}(1997)Bougher, Alexander, \&
  Mayr}]{bougher1997upper}
Bougher, S., Alexander, M., \& Mayr, H. 1997, \bibinfo{title}{Upper atmosphere
  dynamics: global circulation and gravity waves,} Venus II: Geology,
  Geophysics, Atmosphere, and Solar Wind Environment, 259

\bibitem[{D.~F. Buscher(2016)Buscher}]{buscher2016simulating}
Buscher, D.~F. 2016, \bibinfo{title}{Simulating large atmospheric phase screens
  using a woofer-tweeter algorithm,} Optics express, 24, 23566

\bibitem[{H. Canovas {et~al.}(2011)Canovas, Rodenhuis, Jeffers, Min, \&
  Keller}]{canovas2011data}
Canovas, H., Rodenhuis, M., Jeffers, S., Min, M., \& Keller, C. 2011,
  \bibinfo{title}{Data-reduction techniques for high-contrast imaging
  polarimetry-Applications to ExPo,} Astronomy \& Astrophysics, 531, A102

\bibitem[{C. Counselman {et~al.}(1980)Counselman, Gourevitch, King, Loriot, \&
  Ginsberg}]{counselman1980zonal}
Counselman, C., Gourevitch, S., King, R., Loriot, G., \& Ginsberg, E. 1980,
  \bibinfo{title}{Zonal and meridional circulation of the lower atmosphere of
  Venus determined by radio interferometry,} Journal of Geophysical Research:
  Space Physics, 85, 8026

\bibitem[{J.~F. {de Haan} {et~al.}(1987){de Haan}, {Bosma}, \&
  {Hovenier}}]{deHaan1987}
{de Haan}, J.~F., {Bosma}, P.~B., \& {Hovenier}, J.~W. 1987,
  \bibinfo{title}{{The adding method for multiple scattering calculations of
  polarized light},} \aap, 183, 371

\bibitem[{W.~A. {de Rooij} \& C.~C.~A.~H. {van der Stap}(1984){de Rooij} \&
  {van der Stap}}]{deRooij1984}
{de Rooij}, W.~A., \& {van der Stap}, C.~C.~A.~H. 1984,
  \bibinfo{title}{{Expansion of Mie scattering matrices in generalized
  spherical functions},} \aap, 131, 237

\bibitem[{J.~J. Goldstein {et~al.}(1991)Goldstein, Mumma, Kostiuk, Deming,
  Espenak, \& Zipoy}]{goldstein1991absolute}
Goldstein, J.~J., Mumma, M.~J., Kostiuk, T., {et~al.} 1991,
  \bibinfo{title}{Absolute wind velocities in the lower thermosphere of Venus
  using infrared heterodyne spectroscopy,} Icarus, 94, 45

\bibitem[{J.~E. {Hansen} \& J.~W. {Hovenier}(1974){Hansen} \&
  {Hovenier}}]{1974JAtS...31.1137H}
{Hansen}, J.~E., \& {Hovenier}, J.~W. 1974, \bibinfo{title}{{Interpretation of
  the polarization of Venus.},} Journal of Atmospheric Sciences, 31, 1137,
  \dodoi{10.1175/1520-0469(1974)031<1137:IOTPOV>2.0.CO;2}

\bibitem[{J.~E. {Hansen} \& L.~D. {Travis}(1974){Hansen} \&
  {Travis}}]{1974Hansen}
{Hansen}, J.~E., \& {Travis}, L.~D. 1974, \bibinfo{title}{{Light scattering in
  planetary atmospheres},} Space Science Reviews, 16, 527

\bibitem[{J. Helbert {et~al.}(2019)Helbert, Vandaele, Marcq, Robert, Ryan,
  Guignan, Rosas-Ortiz, Neefs, Thomas, Arnold, {et~al.}}]{helbert2019venspec}
Helbert, J., Vandaele, A.~C., Marcq, E., {et~al.} 2019, \bibinfo{title}{The
  VenSpec suite on the ESA EnVision mission to Venus,} in Infrared remote
  sensing and instrumentation XXVII, Vol. 11128, SPIE, 18--32

\bibitem[{D.~M. Hunten {et~al.}(2022)Hunten, Colin, Donahue, \&
  Moroz}]{hunten2022venus}
Hunten, D.~M., Colin, L., Donahue, T.~M., \& Moroz, V.~I. 2022, Venus
  (University of Arizona Press)

\bibitem[{N.~I. {Ignatiev} {et~al.}(2009){Ignatiev}, {Titov}, {Piccioni},
  {Drossart}, {Markiewicz}, {Cottini}, {Roatsch}, {Almeida}, \&
  {Manoel}}]{Ignatiev2009}
{Ignatiev}, N.~I., {Titov}, D.~V., {Piccioni}, G., {et~al.} 2009,
  \bibinfo{title}{{Altimetry of the Venus cloud tops from the Venus Express
  observations},} Journal of Geophysical Research (Planets), 114, 0,
  \dodoi{10.1029/2008JE003320}

\bibitem[{S. Jeffers {et~al.}(2012)Jeffers, Min, Waters, Canovas, Rodenhuis,
  de~Juan~Ovelar, Chies-Santos, \& Keller}]{jeffers2012direct}
Jeffers, S., Min, M., Waters, L., {et~al.} 2012, \bibinfo{title}{Direct imaging
  of a massive dust cloud around R Coronae Borealis,} Astronomy \&
  Astrophysics, 539, A56

\bibitem[{W. Kasprzak {et~al.}(1988)Kasprzak, Hedin, Mayr, \&
  Niemann}]{kasprzak1988wavelike}
Kasprzak, W., Hedin, A., Mayr, H., \& Niemann, H. 1988,
  \bibinfo{title}{Wavelike perturbations observed in the neutral thermosphere
  of Venus,} Journal of Geophysical Research: Space Physics, 93, 11237

\bibitem[{K. Kawabata {et~al.}(1980)Kawabata, Coffeen, Hansen, Lane, Sato, \&
  Travis}]{kawabata1980cloud}
Kawabata, K., Coffeen, D., Hansen, J., {et~al.} 1980, \bibinfo{title}{Cloud and
  haze properties from Pioneer Venus polarimetry,} Journal of Geophysical
  Research: Space Physics, 85, 8129

\bibitem[{J.~C. {Kemp} {et~al.}(1987){Kemp}, {Henson}, {Steiner}, \&
  {Powell}}]{1987kemp}
{Kemp}, J.~C., {Henson}, G.~D., {Steiner}, C.~T., \& {Powell}, E.~R. 1987,
  \bibinfo{title}{{The optical polarization of the sun measured at a
  sensitivity of parts in ten million},} Nature, 326, 270

\bibitem[{T. Kitahara {et~al.}(2019)Kitahara, Imamura, Sato, Yamazaki, Lee,
  Yamada, Watanabe, Taguchi, Fukuhara, Kouyama,
  {et~al.}}]{kitahara2019mountain}
Kitahara, T., Imamura, T., Sato, T.~M., {et~al.} 2019,
  \bibinfo{title}{Stationary features at the cloud top of Venus observed by
  Ultraviolet Imager onboard Akatsuki,} Journal of Geophysical Research:
  Planets, 124, 1266

\bibitem[{O. {Korablev} {et~al.}(2012){Korablev}, {Fedorova}, {Bertaux},
  {Stepanov}, {Kiselev}, {Kalinnikov}, {Titov}, {Montmessin}, {Dubois},
  {Villard}, {Sarago}, {Belyaev}, {Reberac}, \& {Neefs}}]{Korablev2012}
{Korablev}, O., {Fedorova}, A., {Bertaux}, J.-L., {et~al.} 2012,
  \bibinfo{title}{{SPICAV IR acousto-optic spectrometer experiment on Venus
  Express},} Planetary and Space Science, 65, 38,
  \dodoi{10.1016/j.pss.2012.01.002}

\bibitem[{Y.~J. Lee {et~al.}(2024)Lee, Munoz, Choi, Michaelis, Grott, Kang,
  Moon, Yoon, Oh, Zubko, {et~al.}}]{lee2024long}
Lee, Y.~J., Munoz, A., Choi, Y.-J., {et~al.} 2024, \bibinfo{title}{Long-term
  Monitoring Plan of Venus using Earth-orbiting CubeSats,}

\bibitem[{S.~S. Limaye \& V.~E. Suomi(1981)Limaye \& Suomi}]{limaye1981cloud}
Limaye, S.~S., \& Suomi, V.~E. 1981, \bibinfo{title}{Cloud motions on Venus:
  Global structure and organization,} Journal of the Atmospheric Sciences, 38,
  1220

\bibitem[{H. Mayr {et~al.}(1988)Mayr, Harris, Kasprzak, Dube, \&
  Varosi}]{mayr1988gravity}
Mayr, H., Harris, I., Kasprzak, W., Dube, M., \& Varosi, F. 1988,
  \bibinfo{title}{Gravity waves in the upper atmosphere of Venus,} Journal of
  Geophysical Research: Space Physics, 93, 11247

\bibitem[{A. Migliorini {et~al.}(2011)Migliorini, Altieri, Zasova, Piccioni,
  Bellucci, Moinelo, Drossart, D’Aversa, Carrozzo, Gondet,
  {et~al.}}]{migliorini2011oxygen}
Migliorini, A., Altieri, F., Zasova, L., {et~al.} 2011, \bibinfo{title}{Oxygen
  airglow emission on Venus and Mars as seen by VIRTIS/VEX and OMEGA/MEX
  imaging spectrometers,} Planetary and Space Science, 59, 981

\bibitem[{M. Min {et~al.}(2013)Min, Jeffers, Canovas, Rodenhuis, Keller, \&
  Waters}]{min2013color}
Min, M., Jeffers, S.~V., Canovas, H., {et~al.} 2013, \bibinfo{title}{The color
  dependent morphology of the post-AGB star HD 161796,} Astronomy \&
  Astrophysics, 554, A15

\bibitem[{R. Moissl {et~al.}(2009)Moissl, Khatuntsev, Limaye, Titov,
  Markiewicz, Ignatiev, Roatsch, Matz, Jaumann, Almeida,
  {et~al.}}]{moissl2009venus}
Moissl, R., Khatuntsev, I., Limaye, S., {et~al.} 2009, \bibinfo{title}{Venus
  cloud top winds from tracking UV features in Venus Monitoring Camera images,}
  Journal of Geophysical Research: Planets, 114

\bibitem[{D. Mortimer(2023)Mortimer}]{mortimer2013}
Mortimer, D. 2023, Telescope Simulator,,
  \url{https://github.com/dmortimer101/Telescope_Simulator} GitHub

\bibitem[{I.~C. M{\"u}ller-Wodarg {et~al.}(2016)M{\"u}ller-Wodarg, Bruinsma,
  Marty, \& Svedhem}]{muller2016situ}
M{\"u}ller-Wodarg, I.~C., Bruinsma, S., Marty, J.-C., \& Svedhem, H. 2016,
  \bibinfo{title}{In situ observations of waves in Venus’s polar lower
  thermosphere with Venus Express aerobraking,} Nature Physics, 12, 767

\bibitem[{E. Neefs {et~al.}(2024)Neefs, Vandaele, De~Cock, Erwin, Robert,
  Thomas, Berkenbosch, Jacobs, Bogaert, Beeckman, {et~al.}}]{neefs2024venspec}
Neefs, E., Vandaele, A.~C., De~Cock, R., {et~al.} 2024,
  \bibinfo{title}{VenSpec-H spectrometer on the ESA EnVision mission: Design,
  modeling, analysis,}

\bibitem[{M. Newman {et~al.}(1984)Newman, Schubert, Kliore, \&
  Patel}]{newman1984zonal}
Newman, M., Schubert, G., Kliore, A.~J., \& Patel, I.~R. 1984,
  \bibinfo{title}{Zonal winds in the middle atmosphere of Venus from Pioneer
  Venus radio occultation data,} Journal of the atmospheric sciences, 41, 1901

\bibitem[{J. Peralta {et~al.}(2017)Peralta, Hueso, S{\'{a}}nchez-Lavega, Lee,
  Mu{\~{n}}oz, Kouyama, Sagawa, Sato, Piccioni, Tellmann, Imamura, \&
  Satoh}]{Peralta2017}
Peralta, J., Hueso, R., S{\'{a}}nchez-Lavega, A., {et~al.} 2017,
  \bibinfo{title}{Stationary waves and slowly moving features in the night
  upper clouds of Venus,} Nature Astronomy, 1, 0187,
  \dodoi{10.1038/s41550-017-0187}

\bibitem[{M. Persson(2015)Persson}]{persson2015venus}
Persson, M. 2015, Venus thermosphere densities as revealed by Venus Express
  torque and accelerometer data,

\bibitem[{A. Piccialli {et~al.}(2014)Piccialli, Titov, Sanchez-Lavega, Peralta,
  Shalygina, Markiewicz, \& Svedhem}]{piccialli2014high}
Piccialli, A., Titov, D.~V., Sanchez-Lavega, A., {et~al.} 2014,
  \bibinfo{title}{High latitude gravity waves at the Venus cloud tops as
  observed by the Venus Monitoring Camera on board Venus Express,} Icarus, 227,
  94

\bibitem[{M. Rodenhuis {et~al.}(2012)Rodenhuis, Canovas, Jeffers,
  de~Juan~Ovelar, Min, Homs, \& Keller}]{rodenhuis2012extreme}
Rodenhuis, M., Canovas, H., Jeffers, S., {et~al.} 2012, \bibinfo{title}{The
  extreme polarimeter: design, performance, first results and upgrades,} in
  Ground-based and Airborne Instrumentation for Astronomy IV, Vol. 8446, SPIE,
  1413--1430

\bibitem[{M. Rodenhuis {et~al.}(2008)Rodenhuis, Canovas, Jeffers, \&
  Keller}]{rodenhuis2008extreme}
Rodenhuis, M., Canovas, H., Jeffers, S., \& Keller, C. 2008,
  \bibinfo{title}{The Extreme Polarimeter (ExPo): design of a sensitive imaging
  polarimeter,} in Ground-based and Airborne Instrumentation for Astronomy II,
  Vol. 7014, International Society for Optics and Photonics, 70146T

\bibitem[{M. {Rodenhuis} \& C.~U. {Keller}(2007){Rodenhuis} \&
  {Keller}}]{2007lyot.confQ..43R}
{Rodenhuis}, M., \& {Keller}, C.~U. 2007, \bibinfo{title}{{Design Options for
  the Extreme Polarimeter (ExPo)},} in Proceedings of the conference In the
  Spirit of Bernard Lyot: The Direct Detection of Planets and Circumstellar
  Disks in the 21st Century. June 04 - 08, 2007. University of California,
  Berkeley, CA, USA. Edited by Paul Kalas., ed. P.~{Kalas}, 43--+

\bibitem[{L. Rossi {et~al.}(2015)Rossi, Marcq, Montmessin, Fedorova, Stam,
  Bertaux, \& Korablev}]{rossi2015preliminary}
Rossi, L., Marcq, E., Montmessin, F., {et~al.} 2015,
  \bibinfo{title}{Preliminary study of Venus cloud layers with polarimetric
  data from SPICAV/VEx,} Planetary and Space Science, 113, 159

\bibitem[{ {Rossi, L.} \&  {Stam, D. M.}(2018){Rossi, L.} \& {Stam, D.
  M.}}]{rossi2018circpol}
{Rossi, L.}, \& {Stam, D. M.} 2018, \bibinfo{title}{Circular polarization
  signals of cloudy (exo)planets,} A\&A, 616, A117,
  \dodoi{10.1051/0004-6361/201832619}

\bibitem[{M. {Sato} {et~al.}(1996){Sato}, {Travis}, \& {Kawabata}}]{Sato1996}
{Sato}, M., {Travis}, L., \& {Kawabata}, K. 1996,
  \bibinfo{title}{{Photopolarimetry Analysis of the Venus Atmosphere in Polar
  Regions},} Icarus, 124, 569, \dodoi{10.1006/icar.1996.0231}

\bibitem[{M. {Sneep} \& W. {Ubachs}(2005){Sneep} \&
  {Ubachs}}]{2005JQSRT..92..293S}
{Sneep}, M., \& {Ubachs}, W. 2005, \bibinfo{title}{{Direct measurement of the
  Rayleigh scattering cross section in various gases},} J. Quant. Spectrosc.
  Radiative Transfer, 92, 293, \dodoi{10.1016/j.jqsrt.2004.07.025}

\bibitem[{D. Stam {et~al.}(1999)Stam, De~Haan, Hovenier, \&
  Stammes}]{stam1999degree}
Stam, D., De~Haan, J., Hovenier, J., \& Stammes, P. 1999,
  \bibinfo{title}{Degree of linear polarization of light emerging from the
  cloudless atmosphere in the oxygen A band,} Journal of Geophysical Research:
  Atmospheres, 104, 16843

\bibitem[{L.~D. {Travis}(1979){Travis}}]{1979SPIE..183..299T}
{Travis}, L.~D. 1979, \bibinfo{title}{{Imaging and polarimetry with the Pioneer
  Venus Orbiter Cloud Photopolarimeter},} in Proceedings SPIE, Vol. 183, Space
  optics, ed. C.~L. {Wyman}, 299--304, \dodoi{10.1117/12.957426}

\bibitem[{L.~D. Travis(1979)Travis}]{Travis1979}
Travis, L.~D. 1979, Imaging And Polarimetry With The Pioneer Venus Orbiter
  Cloud Photopolarimeter, \dodoi{10.1117/12.957426}

\end{thebibliography}
\bibliographystyle{aasjournalv7}

\end{document}